\documentclass[reprint,amsmath,amssymb,aps,pra,nofootinbib]{revtex4-2}
\usepackage{graphicx}
\usepackage{appendix}
\usepackage{xcolor}
\usepackage[]{hyperref}
\def\mytitle{Scattering solution of interacting Hamiltonian for electronic control of molecular spin qubits}

\newcommand{\myeq}[2]{\cite[Eq. (#2) of Ref. ][]{#1}}
\renewcommand{\myeq}[2]{\cite{#1}}

\begin{document}

\preprint{APS/123-QED}

\author{Christian Bunker$^1$}
\author{Silas Hoffman$^1$}\email{silas.hoffman@ufl.edu}
\author{Jie-Xiang Yu$^2$}
\author{Xiao-Guang Zhang$^1$}
\author{Hai-Ping Cheng$^1$}\email{hping@ufl.edu}\affiliation{$^1$Department of Physics, Center for Molecular Magnetic Quantum Materials, and Quantum Theory Project, University of Florida, Gainesville, Florida 32611, USA \\ $^2$School of Physical Science and Technology, Soochow University, Suzhou, China}
\title{\mytitle}
\date{\today}

\begin{abstract}
We theoretically study how a scattered electron can entangle molecular spin qubits (MSQs). This requires solving the inelastic transport of a single electron through a scattering region described by a tight-binding interacting Hamiltonian. We accomplish this using a Green's function solution. We can model realistic physical implementations of MSQs by parameterizing the tight-binding Hamiltonian with first-principles descriptions of magnetic anisotropy and exchange interactions. We find that for two-MSQ systems with inversion symmetry, the spin degree of freedom of the scattered electron offers probabilistic control of the degree of entanglement between the MSQs.
\end{abstract}

\maketitle


\section{Introduction}
\label{sec:introduction}

Any platform for quantum information processing (QIP) must support entanglement between qubits to achieve quantum speed-up~\cite{nielsen, div_qis, castagnoli}. Molecular spin qubits (MSQs) formed from a two-level subspace of the electron spin degrees of freedom of a molecular system~\cite{sessoli} are a promising platform for QIP because they can be chemically tailored to achieve desired behavior~\cite{sessoli_van, zadrozny} and appear well suited for deploying at scale~\cite{hill}. Although MSQs can be entangled via a Heisenberg-like exchange interaction~\cite{div_loss}, controlling the degree of entanglement experimentally by switching the interaction on and off presents a distinct challenge~\cite{hill}. It would therefore be beneficial to engineer an alternative method to control of the degree of entanglement between MSQs.

A promising method from the solid state qubit community is to use an ancillary electron to mediate the entanglement. In one proposal, a localized ancillary electron has an exchange interaction with two qubits for a certain time interval before being removed~\cite{costa, switzer, switzer2}. Control of the time interval then allows the experimenter to control the degree of entanglement between the qubits~\cite{switzer, switzer2} without needing to switch the Heisenberg-like exchange on and off. However, managing the precise time intervals involved remains experimentally challenging~\cite{costa}.

Alternatively, a delocalized ancillary electron (DAE), sourced by a metallic reservoir, can scatter from two qubits to entangle them \cite{costa}. This allows the degrees of freedom of the DAE itself to control the degree of entanglement between the qubits, replacing the need for precise time intervals \cite{costa, yuasa}. In order to differentiate the proposal of using a \textit{localized} ancillary electron versus a \textit{delocalized} ancillary electron to mediate entanglement, we refer to the latter approach as the `scattering proposal.' In the scattering proposal, schemes for generating Bell states~\cite{ciccarello, ciccarello2} and implementing two-qubit gates~\cite{maruri, ciccarello_gate} have been theoretically demonstrated. However, because the magnetic anisotropy and Heisenberg-like exchange interactions present in MSQs enable inelastic scattering, addressing MSQs within the scattering proposal presents a distinct theoretical challenge. It therefore remains unclear whether a DAE could entangle MSQs.

Upon developing theoretical tools to overcome this challenge, in this paper we demonstrate that a DAE can mediate entanglement within the scalable, tunable platform offered by MSQs.
In section~\ref{sec:method}, we outline our Green's function solution for scattering from a tight-binding interacting Hamiltonian. In section~\ref{sec:results} we apply this solution to demonstrate a scattering process which generates a Bell state from two initially unentangled MSQs. We enumerate the conditions necessary for this process and show that the incoming kinetic energy of the DAE provides a convenient degree of freedom for maximizing the probability of its occurrence. We then demonstrate a scheme by which the DAE controls the degree of entanglement between two MSQs. Applying these results to a realistic physical implementation of two MSQs, we explore the molecular characteristics best suited for generating Bell states and controlling the degree of entanglement before discussing several real molecular systems which could be used to implement these proposals.

\section{Method}
\label{sec:method}

We now formulate a Green's function solution to the problem of a DAE scattering from two spin-$s$ particles with which it can interact. It is prevalent in the literature~\cite{yuasa,ciccarello,ciccarello2,maruri,ciccarello_gate} to solve these types of problems with a wavefunction matching approach in continuous space following Ref. \cite{menezes}. This is only feasible with analytically solvable scattering potentials. In contrast, the tight-binding approach of Ref.~\cite{khomyakov} can be connected to first-principles calculations done with atomic orbital basis sets and can be implemented numerically in order to handle arbitrarily complicated systems. We adopt this approach in order to consider realistic physical implementations of MSQs.

\begin{figure}
    \centering
    \includegraphics[width =1.0\linewidth]{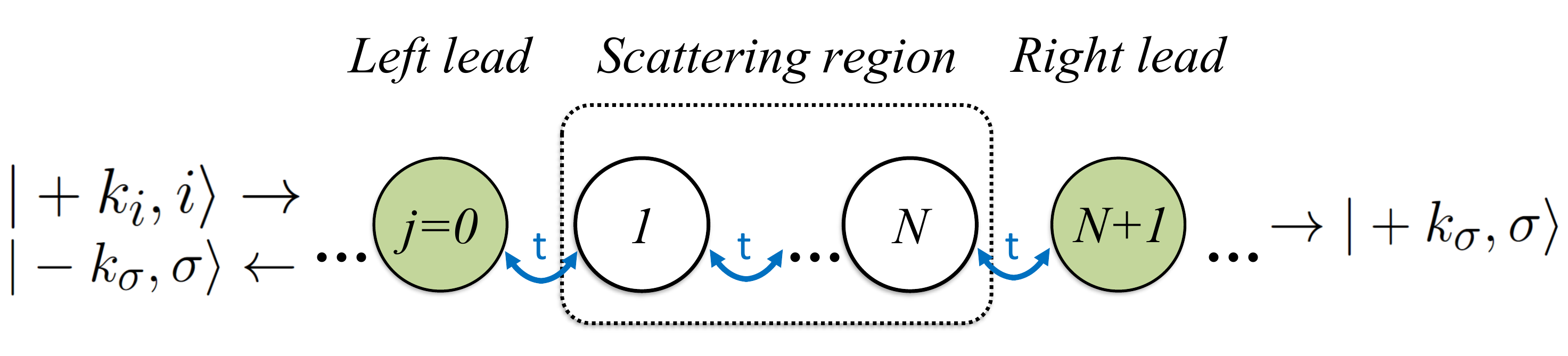}
    \caption{Setup of the one-dimensional scattering problem as an infinite tight-binding chain with nearest neighbor hopping.}
    \label{fig:setup}
\end{figure}

Our scattering setup, sketched in Fig.~\ref{fig:setup}, consists of a single DAE in a one-dimensional wire discretized into sites $j \in \mathbb{Z}$ separated by lattice spacing $a$. These sites form a complete spatial basis $|j\rangle$. The wire could be realized by a single wall carbon nanotube (SWCNT)~\cite{tans} or a silicon nanowire~\cite{serra} exhibiting ballistic transport. The left (right) lead is a noninteracting region of the wire consisting of identical sites $j\leq 0$ ($j>N$) where the DAE wavefunction is a plane wave. The scattering region consists of sites $j=1,... N$ where the DAE wavefunction is no longer a plane wave due to interactions with other particles and external potentials. Although the wire is infinite for our purposes, in practice it would eventually contact a metallic system on either side, as in SWCNT spin valve devices \cite{wernsdorfer, baumgartner}.

Our setup is described by the infinite-dimensional tight-binding Hamiltonian
\begin{flalign}
    \hat{H} =& \sum_{j=-\infty}^\infty \left( \boldsymbol{\varepsilon}_j |j\rangle\langle j| -\textbf{t} |j\rangle \langle j-1| -\textbf{t}|j-1\rangle\langle j| \right) \,.
    \label{eq:Hinfinite}
\end{flalign}
Here boldfaced operators act on three-particle spin states $|\sigma\rangle$, while operators with hats act on both $|\sigma\rangle$ and spatial states $|j\rangle$. Specifically, each $\boldsymbol{\varepsilon}_j$ is an operator in spin space describing the spin physics and on-site energy of site $j$. Because the sites in the leads are all identical, $\boldsymbol{\varepsilon}_{j}= \boldsymbol{\varepsilon}_0$ for $j\leq 0$ or $j>N$.

We are interested in making Eq.~(\ref{eq:Hinfinite}) finite-dimensional. This can be done by taking advantage of the periodicity of the leads. Consider the semi-infinite, periodic Hamiltonian 
\begin{flalign}
    \hat{H}^{(M)} = \sum_{j=-\infty}^{M} \left(  \boldsymbol{\varepsilon}_0 |j \rangle \langle j| - \textbf{t}|j \rangle \langle j-1| - \textbf{t} |j-1 \rangle \langle j| \right) \,.
\end{flalign}
Clearly $\hat{H}^{(0)}$ describes the left lead. We can associate with $\hat{H}^{(M)}$ a Green's function $\hat{g}^{(M)}=(E\hat{I}-\hat{H}^{(M)})^{-1}$ and a surface Green's function~\cite{soven, butler}
\begin{flalign}
    \textbf{g}^{(M)}_{M,M} = \langle M | (E\hat{I}-\hat{H}^{(M)})^{-1} |M\rangle \,.
\end{flalign}

Because $\hat{H}^{(M+1)}$ can be constructed by adding a single site to $\hat{H}^{(M)}$, its Green's function $\hat{g}^{(M+1)}$ satisfies

\begin{flalign}
    \begin{pmatrix}
    E\hat{I} - \hat{H}^{(M)} & \hat{t}^\dagger \\
    \hat{t} & E\textbf{I} - \boldsymbol{\varepsilon}_0 \\
    \end{pmatrix}
    \hat{g}^{(M+1)}
    = \begin{pmatrix}
    \hat{I} & 0 \\
    0 & \textbf{I}
    \end{pmatrix} & \,.
    \label{eq:gf_matrix} 
\end{flalign}
where $\hat{t} = \textbf{t}|M+1\rangle \langle M|$. Solving Eq.~(\ref{eq:gf_matrix}), we obtain \myeq{khomyakov}{23}
\begin{flalign}
    \left( E \textbf{I}-\boldsymbol{\varepsilon}_0 - \textbf{t} \textbf{g}^{(M)}_{M,M} \textbf{t} \right) \textbf{g}^{(M+1)}_{M+1,M+1} = \textbf{I}\,.
\end{flalign}
At the same time, the periodicity of $\hat{H}^{(M)}$ means that $\hat{H}^{(M)} = \hat{H}^{(M+1)}$. As a result, $\textbf{g}^{(M)}_{M,M}$ and  $\textbf{g}^{(M+1)}_{M+1,M+1}$ represent the same quantity, the surface Green's function of a semi-infinite lead. In particular, the surface Green's function of the left lead $\textbf{g}_{00}^{(0)} \equiv \textbf{g}_L$ obeys the self consistency condition \myeq{khomyakov}{24}
\begin{equation}
  \left( E \textbf{I} - \boldsymbol{\varepsilon}_0 - \textbf{t} \textbf{g}_{L} \textbf{t} \right) \textbf{g}_{L} = \textbf{I} \,,\label{eq:gf_selfconsistent}
\end{equation}
which in general can be solved iteratively \cite{haydock, mackinnon}.

We now choose as our basis the eigenstates of the system when the DAE is not interacting with any particles in the scattering region. In this basis, $\boldsymbol{\varepsilon}_0$ is diagonal. Ignoring spin-orbit effects, the nearest neighbor hopping will be spin independent, so $\textbf{t}=t\textbf{I}$. As a result, we can solve for the diagonal elements of Eq.~(\ref{eq:gf_selfconsistent}) \myeq{khomyakov}{79}:
\begin{equation}
    g_{L \sigma \sigma} = \frac{1}{-t} \left( \frac{E-\varepsilon_{0\sigma \sigma}}{-2t} \pm \sqrt{\left(\frac{E-\varepsilon_{0 \sigma \sigma}}{-2t}\right)^2-1} \right)\,.
    \label{eq:gf_surface}
\end{equation}
The sign of the square root is chosen so that the sign of $\text{Im}(g_{L \sigma \sigma})$ is negative, corresponding to the retarded surface Green's function. 

We now introduce the lead self energies $\boldsymbol{\Sigma}_{L} = \textbf{t} \textbf{g}_{L} \textbf{t} = \boldsymbol{\Sigma}_R$ which are equal due to th inversion symmetry of the leads. Substituting Eq.~(\ref{eq:gf_surface}), we obtain
\begin{equation}
    \Sigma_{L \sigma \sigma} = -t \left( \frac{E-\varepsilon_{0 \sigma \sigma}}{-2t} \pm \sqrt{\left(\frac{E-\varepsilon_{0 \sigma \sigma}}{-2t}\right)^2-1} \right)\,.
    \label{eq:selfenergy}
\end{equation}
The left lead self energy, being retarded, encodes an outgoing state, the reflected state. Likewise, the retarded right lead self energy encodes the transmitted state. 

Using the definition of the self energy, Eq.~(\ref{eq:gf_selfconsistent}) becomes
\begin{equation}
    \textbf{g}_{L} = \left[ E \textbf{I} - (\boldsymbol{\varepsilon}_0 + \mathbf{\Sigma}_L) \right]^{-1} \,,
    \label{eq:gf_selfenergy}
\end{equation}
which shows that from the point of view of the Green's function, the physics of the entire left lead can be compactly represented by an energy dependent potential $\boldsymbol{\Sigma}_L$ at its the surface. An analogous expression holds for the right lead. Thus without loss of generality, the entire system can be described by an effective Hamiltonian \myeq{khomyakov}{22}
\begin{flalign}
    \hat{H}' =& \sum_{j=1}^{N+1} \left( \boldsymbol{\varepsilon}_j |j\rangle\langle j| -\textbf{t}|j\rangle \langle j-1| -\textbf{t}|j-1\rangle\langle j| ) \right) \notag \\
    &+(\boldsymbol{\varepsilon}_0+ \boldsymbol{\Sigma}_L) |0\rangle\langle0|+ \boldsymbol{\Sigma}_R |N+1\rangle\langle N+1|\,, 
    \label{eq:Hfinite}
\end{flalign}
so we have succeeded in making Eq.~(\ref{eq:Hinfinite}) finite-dimensional. The corresponding retarded Green's function has elements \myeq{khomyakov}{21} 
\begin{equation}
    \hat{G}_{jj' \sigma \sigma'} = (E\hat{I} - \hat{H}')^{-1}_{jj' \sigma \sigma'} \,.
    \label{eq:gf_elements}
\end{equation} 
Eq.~(\ref{eq:gf_elements}) formally solves the scattering problem because the scattering region wavefunction coefficients
\begin{equation}
        |\psi\rangle = \sum_{j=0}^{N+1} \boldsymbol{\psi}_j |j\rangle = \sum_{j=0}^{N+1} \sum_\sigma \psi_{j \sigma} |j\rangle |\sigma\rangle 
        \label{eq:sr_state}
\end{equation}
can be generated by a convolution \myeq{khomyakov}{19}
\begin{equation}
    \psi_{j \sigma} = \sum_{j' =0}^{N+1} \sum_{\sigma'} G_{jj' \sigma \sigma'} Q_{j' \sigma'}. 
    \label{eq:gf_solution}
\end{equation}
In Eq.~(\ref{eq:gf_solution}), the retarded Green's function encodes the reflected and transmitted states via the retarded self energies. The source vector $\textbf{Q}_{j}=\sum_\sigma Q_{j \sigma}|\sigma\rangle$ encodes the incoming state, which we now discuss in more detail.

The incoming state is defined by both the three-particle spin state $|\sigma\rangle =|i\rangle$ and the spatial wavefunction of the DAE as it impinges on site $j=0$. The spatial wavefunction is a plane wave with wavenumber $+k_{i}$ where $+$ ($-$) indicates a right (left) moving state. We therefore denote the incoming state as $|+k_{i}, i \rangle$. 

The total energy of the system, $E$, is always conserved. In general, $E=\varepsilon_{0\sigma \sigma}+K_\sigma$ where $\varepsilon_{0\sigma\sigma}$ is the potential energy of the three-particle spin state $|\sigma\rangle$ and $K_\sigma$ is the kinetic energy of the DAE plane wave with wavenumber $k_\sigma$, given by the tight-binding dispersion relation
\begin{flalign}
    K_\sigma = 2t -2t\cos(k_\sigma a).
    \label{eq:dispersion}
\end{flalign}
In practice, the experimenter determines $K_i$ by setting the chemical potential of the metallic reservoir that sources the DAE. Without loss of generality, we can choose $\varepsilon_{0 ii} = 0$, so $K_i = E =$
$
    \varepsilon_{0 \sigma \sigma}+2t - 2t \cos(k_{ \sigma} a).
$
This equality determines the wavenumber and velocity of all incoming and outgoing spin states:
\begin{flalign}
    k_{\sigma} =& \frac{1}{a} \cos^{-1} \left( \frac{K_i - \varepsilon_{0 \sigma \sigma}-2t}{-2t} \right)\,
    \label{eq:wavenumber} \\
    v_{ \sigma} =& \frac{1}{\hbar} \frac{d}{dk_{ \sigma}} K_\sigma = \frac{2ta}{\hbar} \sin (k_{ \sigma} a)\,.
    \label{eq:velocity}
\end{flalign}
Note that inelastic scattering occurs when the interactions in the scattering region connect states with different wavenumbers. These wavenumbers and velocities are well defined in the leads because we work in a basis that diagonalizes $\boldsymbol{\varepsilon}_0$; this is the physical reason for choosing such a basis. However, they are not well defined in the scattering region.

Using Eq.~(\ref{eq:gf_solution}), we can match the scattering region wavefunction to the plane waves in the lead. The incoming state is a plane wave
\begin{equation}
    | +k_{ i},i \rangle = \sum_{j=-\infty}^0  A_{\sigma} e^{ik_{ \sigma} ja}|j \rangle |\sigma \rangle \,
    \label{eq:i_state}\,.
\end{equation}
Here $A_\sigma = A \delta_{\sigma i}$, so $A$ specifies the incoming particle amplitude and $\delta_{\sigma i}$ the incoming spin state.  We now introduce the source vector, derived in Eq.~(\ref{A:source_vector}):
\begin{equation}
    Q_{j \sigma} \equiv \frac{i \hbar}{a} A_{\sigma} v_{ \sigma} \delta_{j0} \,
    \label{eq:source_vector}
\end{equation}
where $\delta_{j0}$ specifies that the incoming particle impinges on the scattering region from site $j=0$.
The outgoing states, also plane waves, are given by
\begin{flalign}
    | -k_{ \sigma}, \sigma \rangle =& \sum_{j=-\infty}^0 B_{\sigma} e^{-ik_{ \sigma} ja}|j \rangle |\sigma \rangle \,,
    \label{eq:r_state}  \\
    | +k_{ \sigma}, \sigma \rangle =& \sum_{j=N+1}^\infty  C_{\sigma} e^{ik_{ \sigma} ja}|j \rangle |\sigma \rangle \,.
    \label{eq:t_state}
\end{flalign}
Here $|\sigma\rangle$ can be any outgoing spin state, and $B_{\sigma}$ and $C_{\sigma}$ specify the reflected and transmitted particle amplitude, respectively, in that state.

For any system, once we have specified the spin operators $\boldsymbol{\varepsilon}_j$, we can calculate $\hat{G}$ through Eqs.~-(\ref{eq:Hfinite})-(\ref{eq:gf_elements}). Once the incoming state is specified through Eq.~(\ref{eq:source_vector}), the scattering problem is solved because the wavefunction coefficients from Eq.~(\ref{eq:gf_solution}) determine the outgoing states given in Eq.~(\ref{eq:r_state}) and (\ref{eq:t_state}). In Appendix~\ref{sec:deriv}, we show how this solution generates spin-resolved transmission and reflection coefficients $T_\sigma$ [Eq.~(\ref{eq:Tcoef})] and $R_\sigma$ [Eq.~(\ref{eq:Rcoef})]. In Appendix~\ref{sec:spindep} we apply this solution to a simple example system and demonstrate some of its unique capabilities.

\section{Results}
\label{sec:results}

\begin{figure}
    \centering
    \includegraphics[width =1\linewidth]{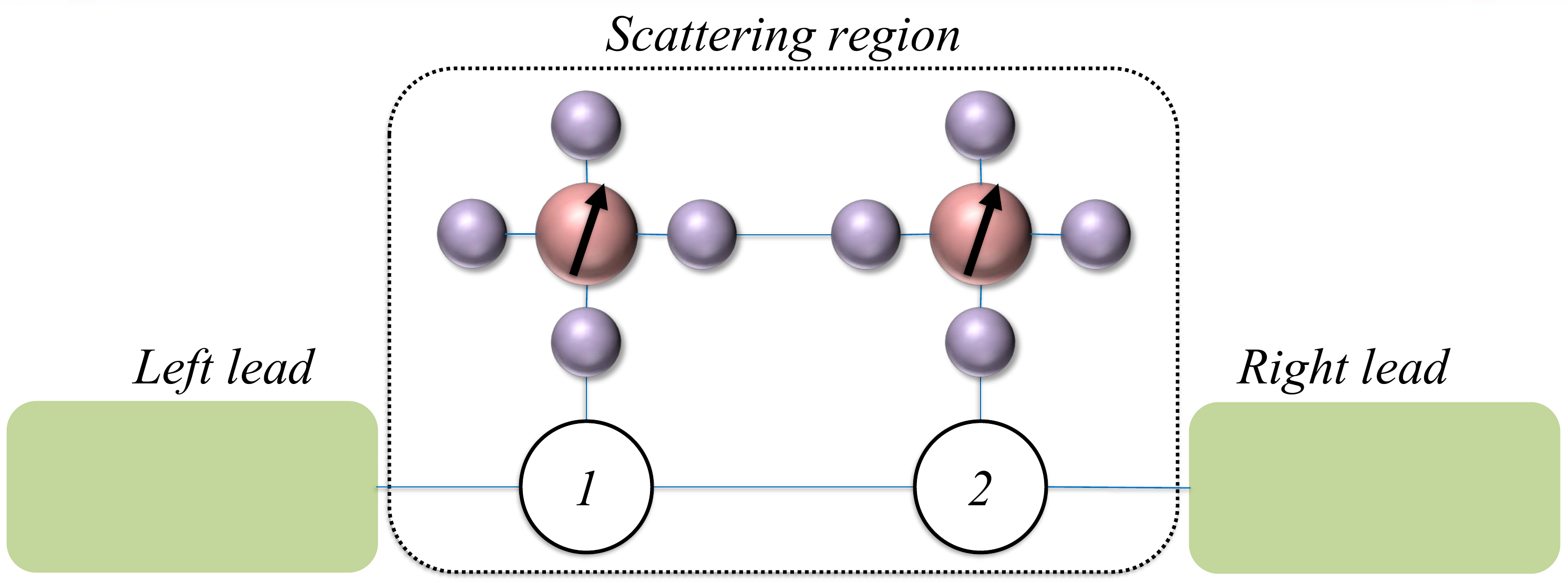}
    \caption{Physical picture of a conduction electron traversing a scattering region of size $N=2$ which models a molecular magnetic system. Blue lines represent nonzero hopping matrix elements. The molecular magnetic system hosts two metal atoms (pink spheres) with electronic spin $s$ (black arrows), which are coupled to each other and to the tight binding sites by ligands (purple spheres).}
    \label{fig:setup_molecule}
\end{figure}

We now apply our solution to a system of a DAE scattering from two spin-$s$ particles. These particles are due to localized electrons in the molecular system forming a composite spin with $2s+1$ levels. The generators of rotations are $S_l^x$, $S_l^y$, and $S_l^z$ where $l=e$ denotes the DAE and $l=1,2$ denotes the spin-$s$ particles. We write these compactly as the vector
\begin{equation}
    \textbf{S}_l = S_l^x \hat{x}+S_l^y \hat{y}+S_l^z\hat{z} \,.
    \label{eq:spinoperator}
\end{equation}

We specify the spin state of the $l^{th}$ particle in terms of the eigenstates $|m_l\rangle_l$ of $S_l^z$, so that three-particle spin states are written $|m_e\rangle_e |m_1\rangle_1 |m_2\rangle_2$. We use $s_{12}$ ($m_{12}$) for the quantum number corresponding to the magnitude ($\hat{z}$ component) of the combined spin operator $\textbf{S}_{12} \equiv \textbf{S}_1 + \textbf{S}_2$ and $s_T$ ($m_T$) for the quantum number corresponding to the magnitude ($\hat{z}$ component) of the total spin operator $\textbf{S}_T \equiv \textbf{S}_e +\textbf{S}_1 + \textbf{S}_2$. We always prepare the system in the three-particle spin state 
\begin{equation}
    |i\rangle = |\text{-}\tfrac{1}{2} \rangle_e |s\rangle_1 |s\rangle_2
    \label{eq:sigmai}
\end{equation}
and we only consider Hamiltonians that are symmetric about the $\hat{z}$-axis in spin space. Therefore, $m_T=2s-1/2$ is conserved and the only accessible three-particle spin states are $|i\rangle$, $|\tfrac{1}{2} \rangle_e |s-1\rangle_1 |s \rangle_2$, and $|\tfrac{1}{2} \rangle_e |s\rangle_1 |s-1\rangle_2$. Inspecting these three states, we see that the $l^{th}$ spin-$s$ particle is restricted to the two-level subspace $\{ |s\rangle_l, |s-1\rangle_l \}$, therefore encoding a MSQ. From the latter two three-particle spin states, we can form two states in which the MSQs are in a Bell state:
\begin{flalign}
    |\pm\rangle =& \, |\tfrac{1}{2} \rangle_e \frac{1}{\sqrt{2}} \left( |s\rangle_1 |s-1\rangle_2 \pm |s-1\rangle_1 |s\rangle_2 \right)\,.
    \label{eq:plusminus} 
\end{flalign}
Since $|i\rangle$,$|+\rangle$, and $|-\rangle$ are all eigenstates of $\textbf{S}_{12}$, with $s_{12}=2s$ for both $|i\rangle$ and $|+\rangle$, processes which conserve $s_{12}$ will be of interest for entangling the MSQs.

As sketched in Fig.~(\ref{fig:setup_molecule}), we have in mind a scattering geometry wherein the DAE can traverse the system without hopping onto the MSQs. Due to the charge of the electrons forming the MSQs, the Coulombic cost of such hopping will be large. Instead, we treat the hopping onto the MSQs perturbatively using a Schreiffer-Wolff transformation \cite{schrieffer} to recover an effectively one-dimensional geometry. Due to this treatment (discussed in more detail in Appendix~\ref{sec:spindep} and Ref. \cite{koch}) the Coulomb interaction between charges occupying the same site becomes an exchange interaction between spins occupying adjacent sites. Therefore in our setup, the first MSQ interacts with site $j=1$ and the second with site $j=N$. This setup allows us to focus purely on the spin-dependent transport effects rather than electronic transport effects such as the Coulomb blockade (see the supplementary information of Ref.~\cite{wernsdorfer_nuc}). Physically, this setup could be achieved by laterally coupling a molecular magnetic system to a SWCNT (see Fig. 1 of Ref. \cite{wernsdorfer}).

To implement this model, we specify the $\boldsymbol{\varepsilon}_j$ operators [which determine the Hamiltonian via Eq.~(\ref{eq:Hfinite})] as
\begin{equation}
    \boldsymbol{\varepsilon}_j = \frac{J}{\hbar^2} \textbf{S}_e \cdot \left(   \textbf{S}_1 \delta_{1j} +  \textbf{S}_2 \delta_{Nj} \right) \,.
    \label{eq:Kondo_cont}
\end{equation}
These operators specify a contact interaction in the sense that the DAE only interacts with each MSQ on a single site. The dot product of spin operators has the same form as the Kondo interaction between conduction electrons and a magnetic impurity in a metal~\cite{schrieffer}. We consider single electron scattering, so there is no Fermi surface and therefore no Kondo physics present in our treatment. However, since this Kondo-like form has been applied to single electron scattering~\cite{costa, ciccarello, ciccarello2, yuasa, maruri, menezes}, we also use this form in order to maintain continuity with previous works. In the following, we consider the effects of scattering from the Kondo-like interaction of Eq.~(\ref{eq:Kondo_cont}) for systems of two simplified MSQs, then two realistic MSQs with appropriate physical symmetries.

\subsection{Two spin-1/2 MSQs}
\label{sec:model12}

We first consider the simplest possible implementation of two MSQs: two spin-1/2 particles which do not interact with each other. Each could be realized by a single electron localized to a molecular orbital with strong $d$ or $f$ character. Analogous mesoscopic solid state systems with magnetic impurities have also been studied in the scattering proposal \cite{costa, ciccarello, ciccarello2, yuasa, maruri, menezes}. The only spin physics present in this system is the Kondo-like interaction between the DAE and the MSQs [Eq.~(\ref{eq:Kondo_cont})]. For this interaction, $[\boldsymbol{\varepsilon}_0, \textbf{S}_{12}^2]=0$. As a result, we work in the eigenbasis of $\textbf{S}_{12}$, $|\sigma\rangle \in \{ |+ \rangle, |- \rangle, |i\rangle \}$, and calculate the corresponding transmission coefficients $T_+$, $T_-$, and $T_i$ using Eq.~(\ref{eq:Tcoef}). Note that in this basis, $\varepsilon_{0\sigma \sigma}$ is the same for all $\sigma$, so the plane wave wavenumbers and velocities given in Eqs.~(\ref{eq:wavenumber}) and (\ref{eq:velocity}) are spin-independent.


In general, the transmission coefficients depend on the DAE's incoming kinetic energy $K_i$, as well as $J$ and $N$ via Eq.~(\ref{eq:Kondo_cont}). More precisely, $t$ sets the energy scale, so $N$, $J/t$, and $K_i/t$ are the free parameters of the problem. We first examine the limit of no spatial separation, in which $N=1$ and Eq.~(\ref{eq:Kondo_cont}) is
\begin{flalign}
    \boldsymbol{\varepsilon}_j =& \frac{J}{\hbar^2} \textbf{S}_e \cdot \textbf{S}_{12} \delta_{1j} \,.
    \label{eq:Kondo_cont_N1}
\end{flalign}
In this limit, the three-particle system reduces to a two-particle system consisting of the DAE and the combined spin $\textbf{S}_{12}$. The scattering then conserves the magnitude of the combined spin $s_{12}$. As noted earlier, this situation is of interest because $|i\rangle$, in which the MSQs are unentangled, can be scattered into $|+\rangle$, in which the MSQs are in a Bell state.

\begin{figure}
    \centering
    \includegraphics[width =\linewidth]{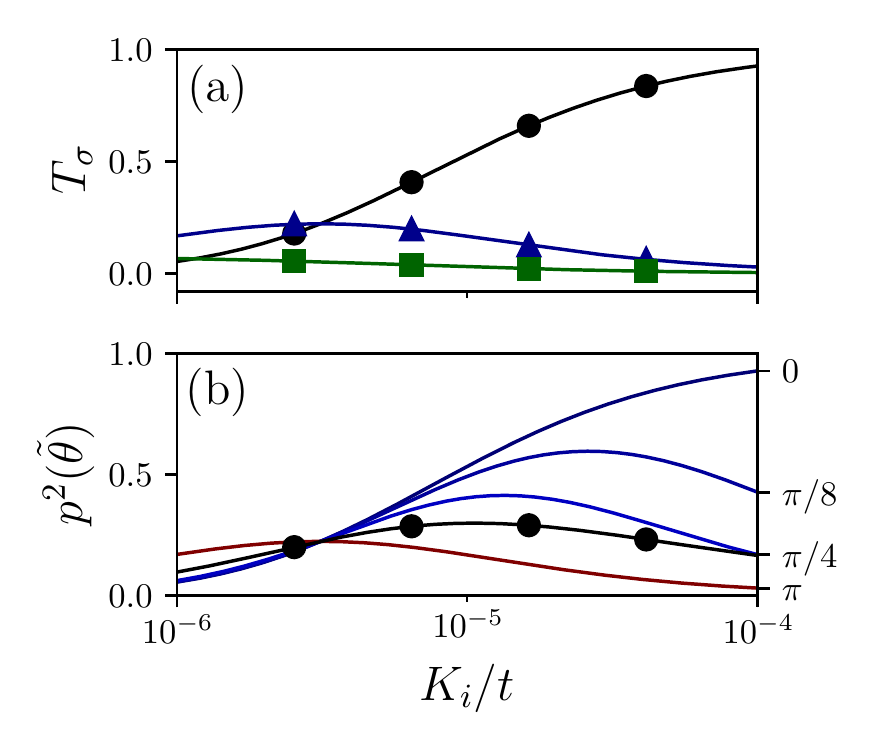}
    \caption{(a) transmission coefficients (black circles) $T_i$, (blue triangles) $T_+$, and (green squares) $ T_- \times 10^5$. The tiny value of $T_-$ shows that the transmission approximately conserves $s_{12}$. (b) probability of success [Eq.~(\ref{eq:prefactor_2})] at different values of $\tilde{\theta}$, labeled on the right. (black circles) the probability of success averaged over $\tilde{\theta}$, given by Eq.~(\ref{eq:prefactor_avg}). Tight-binding parameters are $N=2$, $t=100$ meV, and $J=-0.5$ meV.}
    \label{fig:model12}
\end{figure}

While the $N=1$ case is useful to build intuition, localized spins in a molecular magnetic system are typically separated by nanoscale distances. Consequently, we restrict ourselves to the case $N=2$ corresponding to a finite distance $a$ between the MSQs. We ask whether $s_{12}$ is conserved for $N=2$ as for $N=1$. In Fig.~\ref{fig:model12}(a), we show numerically that for $N=2$, $T_-$ is highly suppressed, so that the the transmission process can be said to approximately conserve $s_{12}$ (although the reflection process may not). Based on this evidence, for the rest of this work we assume that when $N=2$, the $s_{12}$ conserving Kondo-like interaction in Eq.~(\ref{eq:Kondo_cont_N1}) well approximates the true, spatially separated Kondo-like interaction in Eq.~(\ref{eq:Kondo_cont}).

With $N=2$ fixed, we examine the effects $J/t$. These can be seen with the help of Ref.~\cite{menezes}, where the problem of scattering from Eq.~(\ref{eq:Kondo_cont_N1}) has been solved in the continuum, $s_{12}=1/2$ case. This solution, discussed in more detail in Appendix \ref{sec:spindep}, applies to our system when $k_i a \ll 1$ and $N=2$, in which case we expect $T_+ \approx T_{f,c}$ [Eq.~(\ref{eq:continuum_Tf})]. $T_{f,c}$ depends on $J/t$ only through the dimensionless quantity $(J/tk_i a)^2$. Therefore if $T_+ \approx T_{f,c}$ holds, the sign of $J$ has no effect and the magnitude of $J$ affects the value of $k_i$ at which $T_+$ peaks but not the amplitude of the peak. We therefore do not investigate different $J$ values but focus on $J=-0.5$ meV as measured for molecules laterally coupled to SWCNTs \cite{wernsdorfer}. 


Restricting ourselves to $N=2$, $J = -0.5$ meV, and $t=100$ meV for the rest of this work, for spin-1/2 MSQs the transmission coefficients depend only on $K_i$. In Fig.~\ref{fig:model12}(a) we see that $K_i$ provides a convenient degree of freedom for maximizing the Bell state generation probability $T_+$ (blue triangles). We observe a peak $T_+ = 0.22$, consistent with the prediction $\text{max}(T_{f,c})=1/4$ made by Ref.~\cite{menezes}.

\begin{figure}[t]
    \centering
    \includegraphics[width =0.9\linewidth]{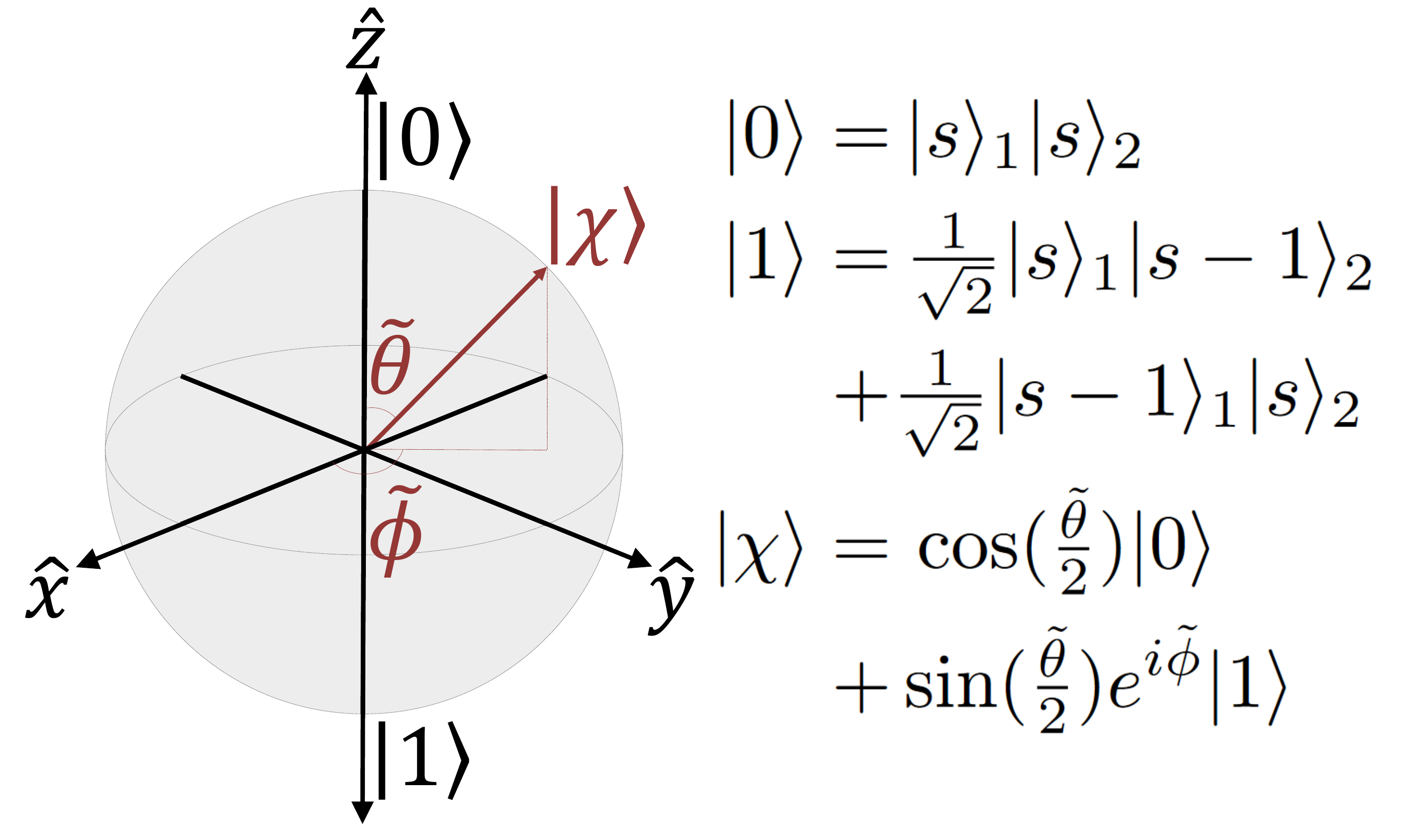}
    \caption{Depiction of the logical state $|\chi\rangle$ in the Bloch sphere formalism. The north pole is completely unentangled, while the south pole is maximally entangled, so specifying $|\chi\rangle$ controls the degree of entanglement.}
    \label{fig:bloch}
\end{figure}

Furthermore, the the DAE's spin degree of the freedom allows us to control the degree of entanglement between MSQs. By control of the degree of entanglement we mean that in the logical basis  
\begin{flalign}
   |0\rangle \equiv& |s\rangle_1 |s\rangle_2\,, \\
   |1\rangle \equiv& \frac{1}{\sqrt{2}} (|s\rangle_1 |s-1\rangle_2 + |s-1\rangle_1 |s\rangle_2)\,,
\end{flalign}
we can rotate from $|0\rangle$ to any desired superposition of $|0\rangle$ and $|1\rangle$ (see Ref.~\cite{switzer2}). This rotation controls the degree of entanglement because $|0\rangle$ is an unentangled state, easily initialized by application of an external magnetic field along the $\hat{z}$ axis, while $|1\rangle$ is a Bell state. These form the antipodal points of a Bloch sphere as shown in Fig.~\ref{fig:bloch}. As we will show, the DAE's spin degree of freedom exactly specifies the desired superposition, leading to control of the degree of entanglement.

When $s_{12}$ is conserved, the transmitted state is
\begin{flalign}
      |T\rangle = \sum_{j=N+1}^\infty \left( C_{i} e^{ik_{i} ja}|j \rangle |i\rangle + C_{+} e^{ik_{+} ja}|j \rangle |+\rangle \right) \,.
    \label{eq:t_state_s12}
\end{flalign}
In Appendix~\ref{sec:deriv}, we show that measuring the observable $\textbf{S}_e \cdot \hat{n}$, where $\hat{n}=(1,\theta,\phi)$ is a unit vector in real space, allows us to project Eq.~(\ref{eq:t_state_s12}) onto the combined spin state
\begin{flalign}
    |\chi \rangle =& \, p(\tilde{\theta}) \left[ \cos(\tfrac{\tilde{\theta}}{2}) |0 \rangle + \sin(\tfrac{\tilde{\theta}}{2}) e^{i\tilde{\phi}}|1\rangle\, \right] \, \label{eq:logical_tilde}
\end{flalign}
The square brackets enclose a unit vector on the Bloch sphere \cite{nielsen}, parameterized by the logical space angles 
\begin{flalign}
\tan(\tfrac{\tilde{\theta}}{2}) =&\, \sqrt{\frac{T_+}{T_i}}\tan(\tfrac{\theta}{2})\,, \label{eq:theta_tilde} \\
\tilde{\phi} =&\,\phi + \phi_+ + \pi\,, \label{eq:phi_tilde}
\end{flalign}
where $\phi_+$ is defined in Eq.~(\ref{eq:phi+}). Then if $T_i, T_+ \neq 0$, $\hat{n}$ specifies $|\chi\rangle$ because we can choose any $\tilde{\theta} \in [0,\pi]$ by appropriate choice of $\theta$ and any $\tilde{\phi} \in[0, 2\pi]$ by appropriate choice of $\phi$. The prefactor $p(\tilde{\theta})$ accounts for the fact that to project onto $|\chi\rangle$, the DAE must be transmitted and measured to have $\textbf{S}_e \cdot \hat{n}=-\hbar/2$. Specifically, the probability of successfully projecting onto $|\chi \rangle$ is
\begin{flalign}
    p^2(\tilde{\theta}) =& \frac{T_i T_+}{T_+ \cos^2 (\tfrac{\tilde{\theta}}{2})+T_i \sin^2 (\tfrac{\tilde{\theta}}{2})} \,.
    \label{eq:prefactor_2}
\end{flalign}
We plot this probability for representative values of $\tilde{\theta}$ in Fig.~\ref{fig:model12}(b). Note that $p^2(\tilde{\theta})=1$ is only possible for $\tilde{\theta}=0$. Since we are interested in preparing states with all values of $\tilde{\theta}$, a useful figure of merit for our scheme by which the DAE controls the degree of entanglement is obtained by averaging $p^2(\tilde{\theta})$ over $\tilde{\theta}$:
\begin{flalign}
    \overline{p^2} = \frac{1}{\pi} \int_0^\pi  \frac{ T_i T_+ \, d\,\tilde{\theta}}{T_+ \cos^2 (\tfrac{\tilde{\theta}}{2})+T_i \sin^2 (\tfrac{\tilde{\theta}}{2})} = \sqrt{T_i T_+} \,.
    \label{eq:prefactor_avg}
\end{flalign}

In Fig.~\ref{fig:model12}(b), we show how $\overline{p^2}$ varies with $K_i$. As with $T_+$, $K_i$ provides a convenient parameter for maximizing $\overline{p^2}$. We observe max$(\overline{p^2})=0.30$ which we can analyze in light of our previous assumption that for $k_i a \ll 1$ and $N=2$, the continuum results of Ref.~\cite{menezes} hold, i.e. $T_{+} \approx T_{f,c}$ [Eq.~(\ref{eq:continuum_Tf})] and $T_{i} \approx T_{nf,c}$ [Eq.~(\ref{eq:continuum_Tnf})]. Inserting these into Eq.~(\ref{eq:prefactor_avg}), the result has a maximum of $\overline{p^2}=0.32$, so the observed maximum is reasonable.

\begin{figure}[t]
    \centering
    \includegraphics[width =\linewidth]{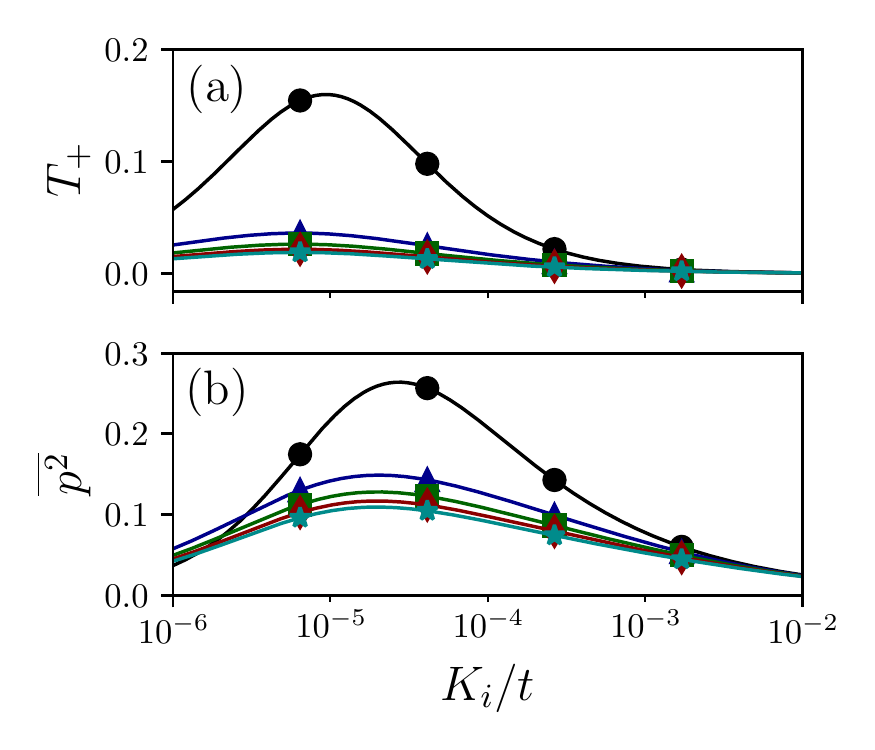}
    \caption{Dependence of (a) $T_+$ and (b) $\overline{p^2}$ on $K_i$ and $\Delta E$ when $s=1$. Tight-binding parameters are $N=2$, $t=100$ meV, $J = -0.5$ meV, and $J_{12}^x = J_{12}^z = 1$ meV. By choice of $D$ we set (black circles) $\Delta E = 0.0$ meV, (blue triangles) $\Delta E = -0.1$ meV, (green squares) $\Delta E = -0.2$ meV, (red diamonds) $\Delta E = -0.3$ meV, and (cyan stars) $\Delta E = -0.4$ meV.}
    \label{fig:model1}
\end{figure}

\subsection{Two molecular magnetic MSQs}

We now consider a more complicated implementation of two MSQs: a molecular magnetic system hosting two metal atoms. Each atom's valence electrons form a composite spin-$s$ particle. As in the previous section, these particles have a Kondo-like interaction with the DAE given by precisely by Eq.~(\ref{eq:Kondo_cont}) and approximately by Eq.~(\ref{eq:Kondo_cont_N1}). In addition, these particles have uniaxial magnetic anisotropy and a Heisenberg-like exchange interaction with each other. In specifying the form of these interactions, we recall that only Hamiltonians which are symmetric about the $\hat{z}$-axis in spin space can encode a MSQ in the subspace $\{ |s\rangle_l, |s-1\rangle_l \}$. To second order in the spin operators, the most general form of $\boldsymbol{\varepsilon}_j$ is
\begin{flalign}
    \boldsymbol{\varepsilon}_j = \frac{1}{\hbar^2} \Big[& J\textbf{S}_e \cdot \textbf{S}_{12} \delta_{j1} 
    + D_1 (S_1^z)^2 + D_2 (S_1^z)^2 \notag \\
    &+ J_{12}^x (S_1^x S_2^x + S_1^y S_2^y) + J_{12}^z S_1^z S_2^z \Big] \,.
    \label{eq:model32}
\end{flalign}
Eq.~(\ref{eq:model32}) is block diagonalized by $m_T$, and we concentrate on the $m_T=2s-1/2$ block. 
In this block, we choose as our basis the eigenbasis of $\textbf{S}_{12}^2$, namely $|\sigma\rangle \in \{ |+\rangle,|-\rangle,|i\rangle \}$. In this basis,
\begin{flalign}
    \boldsymbol{\varepsilon}_j =
    &[2s^2 D + (s^2 - s)J_{12}^z] \textbf{I} \notag \\
    + &\begin{pmatrix} 
    (1-2s)D +sJ_{12}^x & (s-\frac{1}{2})\Delta_D & 0 \\
    (s-\frac{1}{2})\Delta_D & (1-2s)D  -sJ_{12}^x & 0 \\
    0 & 0 & s J_{12}^z  \\
    \end{pmatrix} \notag \\
    +  J &\begin{pmatrix} 
    (s-\frac{1}{2}) & 0 & \sqrt{s} \\ 0 & (s-\frac{1}{2}) & 0 \\ \sqrt{s} & 0& -s
    \end{pmatrix} \delta_{j1}\,, 
    \label{eq:model32_block}
\end{flalign}
where $D=(D_1+D_2)/2$ and $\Delta_D = D_1 - D_2$. The parameters $D_1$, $D_2$, $J_{12}^x$, $J_{12}^y$, and $J_{12}^z$ in Eq.~(\ref{eq:model32}) can be fit with density functional theory (DFT) in order to build a model of a molecular system containing spins with a shared axis of uniaxial magnetic anisotropy; see for example Ref.~\cite{yu}.

\begin{figure}[t]
    \centering
    \includegraphics[width =\linewidth]{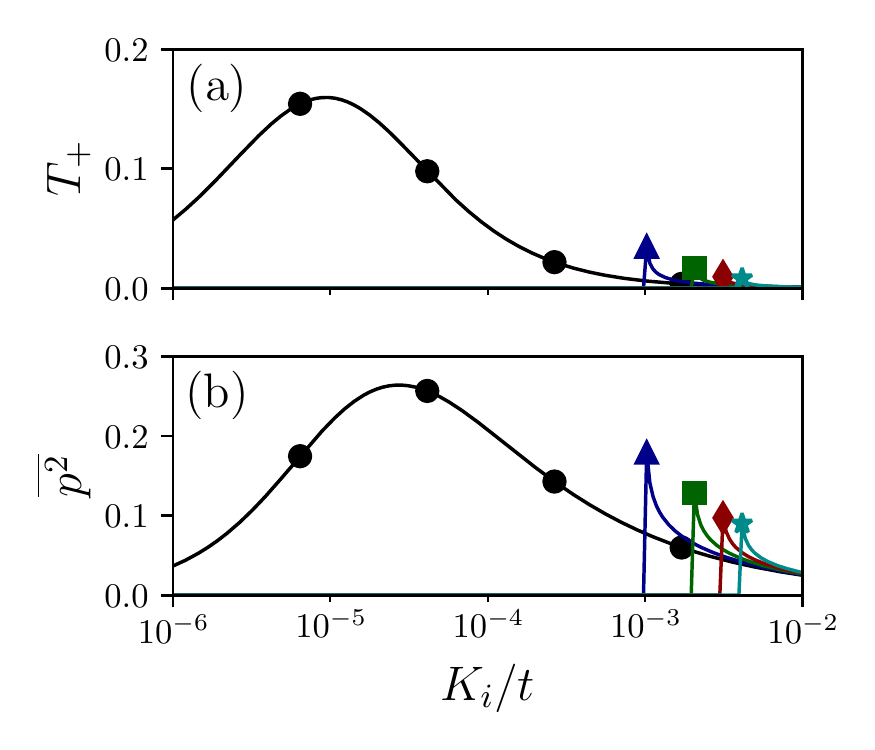}
    \caption{Dependence of (a) $T_+$ and (b) $\overline{p^2}$ on $K_i$ and $\Delta E$ when $s=1$. Tight-binding parameters are $N=2$, $t=100$ meV, $J = -0.5$ meV, and $J_{12}^x = J_{12}^z = 1$ meV. By choice of $D$ we set (black circles) $\Delta E = 0.0$ meV, (blue triangles) $\Delta E = 0.1$ meV, (green squares) $\Delta E = 0.2$ meV, (red diamonds) $\Delta E = 0.3$ meV, and (cyan stars) $\Delta E = 0.4$ meV. }
    \label{fig:model1_positive}
\end{figure}

Since the two particles have the same $s$, the system has inversion symmetry if $D_1 = D_2$. For the rest of this section, we impose inversion symmetry. This is necessary to make $\boldsymbol{\varepsilon}_0$ diagonal so that $s_{12}$ a good quantum number. Then because the Kondo-like interaction conserves $s_{12}$, $|i\rangle$ can only scatter into itself or $|+\rangle$ as before. However, these two states are no longer degenerate; instead, they are split in energy by
\begin{equation}
    \Delta E \equiv \varepsilon_{0++} - \varepsilon_{0ii} = (1-2s)D + s(J_{12}^x - J_{12}^z)\,.
    \label{eq:DeltaE}
\end{equation} 
Inspecting Eq.~(\ref{eq:model32_block}) when $\Delta_D = 0$, we see that aside from $J$, which we have already fixed, the only free parameters that affect the transmission coefficients are $K_i$, $\Delta E$, and $s$. Recalling that the figure of merit for our scheme by which the DAE controls the degree of entanglement is $\overline{p^2}$, we now explore how each of these affect $\overline{p^2}$.

In the $s=1/2$ case (Fig.~\ref{fig:model12}) we saw that $\overline{p^2}$ increases with increasing $K_i$, reaches a maximum, then decreases. This is also the case for the molecular magnetic system when $\Delta E \geq 0$ as shown in Fig.~\ref{fig:model1}. When $\Delta E < 0$, the behavior is very different because transmission into the $|+\rangle$ state is energetically forbidden when $K_i < \Delta E$, as shown in Fig.~\ref{fig:model1_positive}. However, in both cases $\overline{p^2}$ has a single maximum over the domain of $K_i$ which we denote max$(\overline{p^2})$. We assume that we can always tune $K_i$ to achieve max$(\overline{p^2})$.

\begin{figure}[t]
    \centering
    \includegraphics[width =\linewidth]{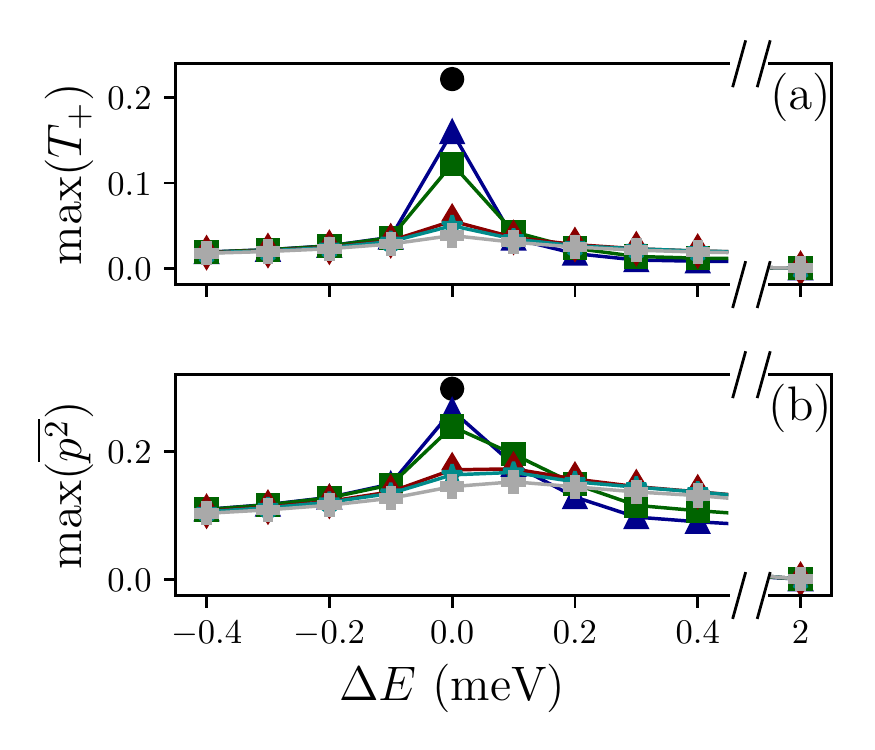}
    \caption{Maxima of (a) $T_+$ and (b) $\overline{p^2}$ for spin-$s$ particles at different $\Delta E$. (black circles) $s=1/2$, for which only $\Delta E/t = 0.0$ is possible, with tight-binding parameters $N=2$, $t=100$ meV, and $J=-0.5$ meV. (blue triangles) $s=1$, (green squares) $s=3/2$, (red diamonds) $s=4$, (cyan stars) $s=9/2$, and (gray pluses) $s=6$, with tight-binding parameters $N=2$, $t=100$ meV, $J=-0.5$ meV, $J_{12}^x = J_{12}^z = 1.0$ meV, and $D$ variable.}
    \label{fig:peaks}
\end{figure}

In Fig.~\ref{fig:peaks}, we explore the dependence of max$(T_+)$ and max$(\overline{p^2})$ on $\Delta E$ and $s$ in order to determine the general molecular characteristics most suitable for generating Bell states and for our scheme by which the DAE controls the degree of entanglement. We plot the $s=1/2$ result max$(\overline{p^2}) = 0.30$ (black circles) for reference. We then plot data for $s=$ 1, 3/2, 4, 9/2 and 6. Note that max$(T_+)$ and max$(\overline{p^2})$ tend to decrease with increasing $s$ consistent with previous results (e.g. Fig. 2 of Ref \cite{ciccarello2}). Also, as $s$ increases towards the classical limit, the dependence of max$(T_+)$ and max$(\overline{p^2})$ on $s$ decreases.

\section{Discussion \& Conclusion}

We showed that molecular magnetic systems hosting two metal atoms with a shared axis of symmetry in spin space and inversion symmetry in real space are suitable for encoding two MSQs and entangling them using a DAE. Specifically, we showed how to generate Bell states and control the degree of entanglement between the MSQs using the spin degree of freedom of the DAE. Although the control scheme we presented is probabilistic, the experimenter immediately sees whether it has succeeded, and we can quantify the probability of success with $\overline{p^2}$ [Eq.~(\ref{eq:prefactor_avg})]. 


\begin{figure}[t]
    \centering
\includegraphics[width =\linewidth]{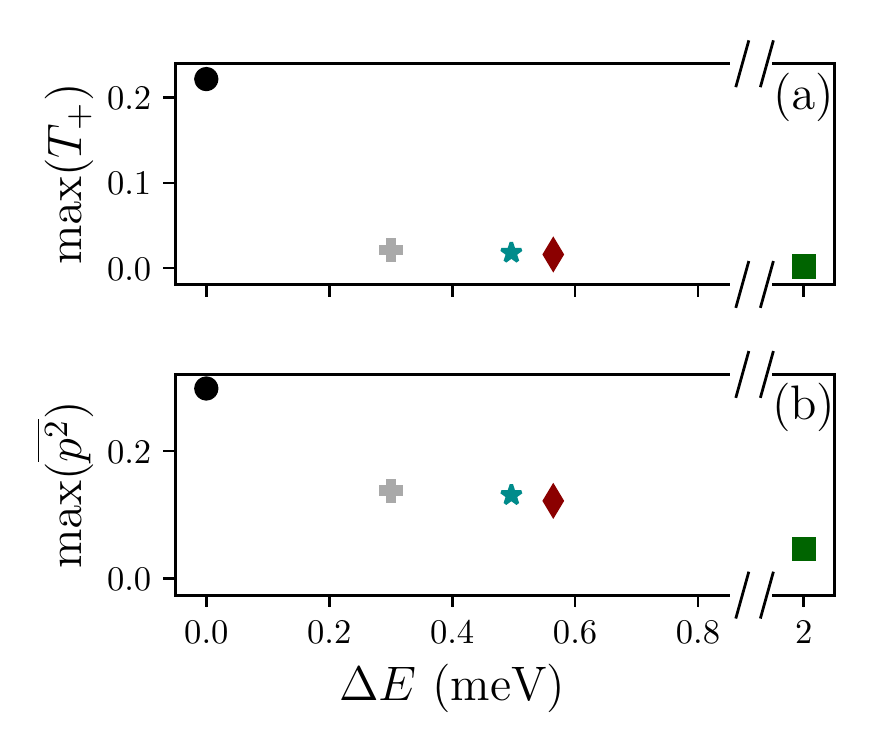}
\caption{Maxima of (a) $T_+$ and (b) $\overline{p^2}$ for real molecular magnetic systems. Tight binding parameters are $N=2$ , $t=100$ meV, and $J=-0.5$ meV throughout. (green square) MnPc, with $s=3/2$, $D=-0.99$ meV, and $J_{12}^x = J_{12}^z = -0.77$ meV \cite{park}. 
(red diamond) Mn(III) dimer, with $s=4$, $D=-0.08$ meV, and $J_{12}^x = J_{12}^z = -0.53$ meV \cite{christou_MnIII}. (cyan star) Mn$_4$ dimer, with $s=9/2$, $D=-0.06$ meV, and $J_{12}^x=J_{12}^z=0.009$ meV \cite{christou_Mn4}. (gray plus) Mn$_3$ dimer, with $s=6$, $D=-0.03$ meV, and $J_{12}^x = J_{12}^y=-0.006 $ meV \cite{yu}.} \label{fig:peaks_real}
\end{figure}

We explored the dependence of $\overline{p^2}$ on $s$, the spin
of the MSQs, and $\Delta E$, the energy splitting. Fig.~\ref{fig:peaks} shows that, under our constraints $N=2$, $t=100$ meV, and $J=-0.5$ meV, systems with $s \leq 3/2$ and $|\Delta E| \ll 1$ meV are best suited for the control scheme we presented. More generally, systems with $|\Delta E/J| \ll 1$  are desirable. If such systems cannot be not found, they could be engineered thanks to the chemical tunability of molecular QIP platforms. DFT has shown that the addition of symmetry-breaking ligands \cite{hooshmand} or charge doping \cite{liu} can lower the magnitude of the magnetic anisotropy of single molecule magnets (SMMs), which would in turn decrease $|\Delta E|$ according to Eq.~(\ref{eq:DeltaE}). Alternatively, one can use external experimental parameters to tune $\Delta E$, e.g. with applied magnetic fields or through the dependence of $J_{12}^x$ on external pressure \cite{yu}.

Our results also motivate discussion of the feasibility of real molecular magnetic systems for implementing our scheme by which the DAE controls the degree of entanglement. To do this, we first recall that the setup we have in mind involves a molecular system laterally coupled to a conducting region which is long and narrow along the $\hat{z}$-axis (e.g. a SWCNT as in Fig. 1 of Ref.~\cite{wernsdorfer}) with the MSQs spatially separated along the $\hat{z}-$axis (see Fig.~\ref{fig:setup_molecule}). A dimer of two identical SMMs held together by a linker naturally fits into our setup because it can be placed onto the SWCNT with the intermolecular axis parallel to the $\hat{z}$-axis. Some examples are
an $s=4$ Mn(III) dimer \cite{christou_MnIII}, an $s=9/2$ Mn$_4$ dimer \cite{christou_Mn4}, and an $s=6$ Mn$_3$ dimer \cite{yu}. All have the requisite symmetries of Eq.~(\ref{eq:model32}) with $D_1 = D_2$ and $|\Delta E|$ of order 0.1 meV as desired. However, as shown in Fig.~\ref{fig:peaks_real} their max$(\overline{p^2})$ values remains well below the $s=1/2$ result due to their large $s$ values. Our results suggest that much improved $\overline{p^2}$ could be achieved by a SMM dimer with $s \leq 3/2$; however, no such complexes have come to our attention.

Metal-phthalocyanines (MPcs) in which organic ligands surround a central metal ion \cite{park} are another real molecular system that offers a potential realization of our scheme. MPcs are already of interest for QIP applications \cite{hadt,wernsdorfer_nuc}. Isolated MPcs are typically planar and could be placed side by side atop the SWCNT to achieve our desired setup. Choice of the metal ion allows us to select a lower spin; for example MnPc has $s=3/2$ \cite{park} while VOPc has $s=1/2$ \cite{hadt}. While the former is hampered by an unusually large energy splitting of $\Delta E = -1.98$ meV, the latter has no energy splitting by virtue of being $s=1/2$ and thus appears especially promising for our scheme by which the DAE controls the degree of entanglement.

Although our Green's function solution achieves increased realism by incorporating first principles descriptions of molecules and accounting for the spatial degrees of freedom of the delocalized electron, it would be interesting to include some additional physical effects for a more complete description. First, our molecular Hamiltonian only accounts for spin degrees of freedom. It would be interesting to incorporate the full orbital degrees of freedom. As demonstrated in Appendix~\ref{sec:physical_origin}, our solution appears capable of this task. Second, we have adopted a simplified picture of the leads as tight-binding chains hosting only a single conduction electron. A full treatment of the leads must recognize the presence of an entire conduction band, leading to Kondo effects.

\section*{Acknowledgements}

We are grateful for helpful discussions with Garnet Chan and Eric Switzer. This work was supported as part of the Center for Molecular Magnetic Quantum Materials, an Energy Frontier Research Center funded by the U.S. Department of Energy, Office of Science, Basic Energy Sciences under Award no. DE-SC0019330. Computations were done using the utilities of the National Energy Research Scientific Computing Center and University of Florida Research Computing.

\appendix

\section{Derivations}
\label{sec:deriv}

\subsection{Source vector}

The role of the retarded Green's function [Eq.~(\ref{eq:gf_elements})] is to connect the incoming state [Eq.~(\ref{eq:i_state})] to the outgoing states [Eqs.~(\ref{eq:r_state})~and~\ref{eq:t_state}]. Although the incoming and outgoing states are boundary conditions in the mathematical sense, it is important to note that only the incoming state can be determined by the experimenter. The outgoing states are determined by the incoming state and the scattering potential. The role of the source vector [Eq.~(\ref{eq:source_vector})] is to specify the incoming state, while the outgoing states are encoded by the retarded Green's function itself.

The eigenstates in the left lead are plane waves specified by $A_\sigma$ and $B_\sigma$, the incoming and reflected particle amplitudes in state $|\sigma\rangle$. By working in the basis that diagonalizes $\boldsymbol{\varepsilon}_0$, we ensure these plane waves have well defined wavenumbers and velocities given by Eqs.~(\ref{eq:wavenumber})~and~(\ref{eq:velocity}). As a result, the wavefunction in the left lead takes the general form
\begin{equation}
    |\psi\rangle = \sum_{j=-\infty}^0 \sum_\sigma \left( A_\sigma e^{ik_{\sigma}ja}+B_\sigma e^{-ik_{\sigma}ja}\right) |j\rangle|\sigma\rangle\,.
    \label{A:LL_state}
\end{equation}
We could obtain a boundary condition at site $j=0$ directly from Eq.~(\ref{A:LL_state}), but it would not properly distinguish the incoming state from the reflected state. Instead, we can use the Schrodinger equation to define a source vector as follows.

Applying Eq.~(\ref{eq:Hinfinite}) to Eq.~(\ref{eq:sr_state}) yields the Schrodinger equation at $j=1$:
\begin{equation}
    (E\textbf{I} - \boldsymbol{\varepsilon}_0) \boldsymbol{\psi}_0 + \textbf{t} \boldsymbol{\psi}_{1}+\textbf{t}\boldsymbol{\psi}_{-1} = 0
    \label{A:schro_spinful}
\end{equation}
where $E$ is the total energy of the system. Assuming that the hopping is spin independent, $\textbf{t}$ is diagonal. Since we already diagonalized $\boldsymbol{\varepsilon}_0$, Eq.~(\ref{A:schro_spinful}) simplifies to
\begin{equation}
    (E - \varepsilon_{0 \sigma \sigma}) \psi_{0\sigma} + t(\psi_{1\sigma}+\psi_{-1\sigma})=0.
    \label{A:schro}
\end{equation}

From Eq.~(\ref{A:LL_state}), we have $\psi_{-1\sigma} = A_\sigma e^{-ik_{\sigma}a}+B_\sigma e^{ik_{\sigma}a} $ and $\psi_{0\sigma} = A_\sigma +B_\sigma$. With these substitutions, Eq. (\ref{A:schro}) can be written as 
\begin{flalign}
    E \psi_{0\sigma} - &\left( \varepsilon_{0 \sigma \sigma} \psi_{0\sigma} - t \psi_{1\sigma} - te^{ik_{\sigma} a} \psi_{0\sigma} \right)\notag \\
    =& A_\sigma t \left( e^{ik_{\sigma} a} - e^{-ik_{\sigma} a} \right)
    \label{A:schro_prime}
\end{flalign}

Following Ref.~\cite{zwierzycki}, Eq.~(\ref{A:schro_prime}) can be rewritten by defining on the left hand side a retarded self energy 
\begin{flalign}
\Sigma_{L\sigma \sigma} =& -t e^{ik_{\sigma} a} \notag \\
=& -t \left( \frac{E-\varepsilon_{0 \sigma \sigma}}{-2t} + \sqrt{\left(\frac{E-\varepsilon_{0 \sigma \sigma}}{-2t}\right)^2 - 1} \right)\,,
\label{A:selfenergy}
\end{flalign}
and on the right hand side, using Eq.~(\ref{eq:velocity}), a source vector
\begin{flalign}
   Q_{0\sigma} =&\, A_\sigma t(e^{ik_{\sigma} a} - e^{-ik_{\sigma} a}) \notag \\  
   =&\, 2iA_\sigma t \sin(k_{\sigma} a)  = \frac{i\hbar }{a} A_\sigma v_{\sigma}\,.
   \label{A:source_vector}
\end{flalign}
Eqs.~(\ref{A:selfenergy})~and~(\ref{A:source_vector}) recover Eqs.~(\ref{eq:selfenergy})~and~(\ref{eq:source_vector}), respectively.
Substituting them yields the Schrodinger equation with an effective Hamiltonian and a source term,
\begin{flalign}
    E \psi_{0\sigma} - &\left[ ( \varepsilon_{0 \sigma \sigma} - \Sigma_{L\sigma\sigma}) \psi_{0\sigma} - t \psi_{1\sigma}  \right] =Q_{0\sigma} \,,
    \label{A:schro_wfm}
\end{flalign}
so that as usual the Green's function, which solves the Schrodinger equation with an identity source, can be convoluted with the source term to solve Eq.~(\ref{A:schro_wfm}).

\subsection{Transmission and reflection coefficients}

Enforcing continuity of Eq.~(\ref{eq:sr_state}) with Eqs.(\ref{eq:i_state}) and~(\ref{eq:r_state}) at $j=0$ and with Eq.~(\ref{eq:t_state}) at $j=N+1$ leads to the boundary conditions
\begin{flalign}
    \psi_{N+1,\sigma} =& \, C_\sigma \label{eq:contLL} \\
    \psi_{0, \sigma} =& \, A_\sigma + B_\sigma \,.
    \label{eq:contRL}
\end{flalign}
We can match the coefficients of each spin state individually because $j=0$ and $j=N+1$ are in the leads, where there are no interactions to couple different spin states.

The transmission (reflection) coefficients can now be calculated from the ratio of transmitted (reflected) flux to incoming flux. The incoming flux is $ \sum_{\sigma'} A_{\sigma'} A_{\sigma'} v_{ \sigma'}$ while the transmitted flux in spin state $|\sigma\rangle$ is $|C_\sigma|^2 v_{ \sigma}$. Inserting Eqs.~(\ref{eq:gf_solution}),~(\ref{eq:source_vector}), and~(\ref{eq:contRL}), we have
\begin{flalign}
    T_{\sigma} =& \frac{|C_{\sigma}|^2 v_{ \sigma}}{\sum_{\sigma'} A_{\sigma'} A_{\sigma'} v_{ \sigma'}} \notag \\
    =& \frac{|\tfrac{i\hbar}{a} \sum_{\sigma''} G_{N+1,0,\sigma,\sigma''} A_{\sigma''} v_{ \sigma''} |^2 v_{ \sigma} }{\sum_{\sigma'} A_{\sigma'} A_{\sigma'} v_{ \sigma'}}     \label{eq:Tcoef_general} \\
    =& \frac{\hbar^2}{a^2} |G_{N+1,0,\sigma,i}|^2 v_{ \sigma} v_{ i}\,.
    \label{eq:Tcoef}
\end{flalign}
Note that Eq.~(\ref{eq:Tcoef_general}) is general while Eq.~(\ref{eq:Tcoef}) for the case of a single incoming spin state, $A_\sigma = A \delta_{\sigma i}$. Similarly, the reflected flux in spin state $|\sigma \rangle$ is $|B_{\sigma}|^2 v_{ \sigma}$, so
using Eqs.~(\ref{eq:gf_solution}),~(\ref{eq:source_vector}),~and~(\ref{eq:contLL}), we have
\begin{flalign}
    R_\sigma =& \frac{|B_{\sigma}|^2 v_{ \sigma}}{\sum_{\sigma'} A_{\sigma'} A_{\sigma'} v_{ \sigma'}} \notag \\
    =& \frac{ \left| \tfrac{i \hbar}{a} \sum_{\sigma'} G_{0,0,\sigma, \sigma''} A_{\sigma''} v_{ \sigma''}-A_{\sigma} \right|^2 v_{ \sigma}}{\sum_{\sigma'} A_{\sigma'} A_{\sigma'} v_{ \sigma'}}     \label{eq:Rcoef_general} \\
    =& \left| \frac{i\hbar}{a}G_{0,0,\sigma, i}v_{ i} - \delta_{\sigma i} \right|^2 \frac{v_{\sigma}}{v_{i}}\,.
    \label{eq:Rcoef}
\end{flalign}
Again, Eq.~(\ref{eq:Rcoef}) is for a single incoming spin state.

\subsection{Measuring the spin of the DAE controls the degree of entanglement}

We isolate the spin degrees of freedom of the transmitted state by projecting Eq.~(\ref{eq:t_state_s12}) onto site $j=N+1$:
\begin{flalign}
     \langle j\text{=}N+1 | T \rangle =& \, e^{ik_{i}a(N+1)}C_{i} |i \rangle + e^{ik_{ +}a(N+1)}C_{+} |+ \rangle \notag \\
    =& \, \sqrt{T_i}  |\text{-}\tfrac{1}{2}\rangle_e |0\rangle + \sqrt{T_+} e^{i \phi_+}|\tfrac{1}{2}\rangle_e |1\rangle \,.
    \label{eq:transmitted_state}
\end{flalign}
Here we substituted $|C_\sigma|=\sqrt{T_\sigma}$ and dropped the overall phase $\exp \boldsymbol{(}ik_{i}a(N+1) +\text{arg}(C_i)\textbf{)}$ but allowed for a complex phase between the two spin states
\begin{flalign}
   &\exp(i\phi_+)= \notag \\
   &\exp \boldsymbol{(}i (k_{+}-k_i)a(N+1)+\text{arg}(C_+)-\text{arg}(C_i) \boldsymbol{)}\,. 
   \label{eq:phi+}
\end{flalign}
Measuring the DAE's spin along the unit vector $\hat{n}=(1,\theta,\phi)$ projects the DAE's state onto one of the eigenstates of $\textbf{S}_e \cdot \hat{n}$, namely \cite{shankar}
\begin{flalign}
\left| \Uparrow \right\rangle_e =& \cos (\tfrac{\theta}{2})|\tfrac{1}{2} \rangle_e + \sin(\tfrac{\theta}{2})e^{i\phi}|\text{-}\tfrac{1}{2}\rangle_e\,, \label{eq:Sn_plus} \\
\left| \Downarrow \right\rangle_e =& -\sin(\tfrac{\theta}{2})|\tfrac{1}{2} \rangle_e + \cos(\tfrac{\theta}{2})e^{i\phi}|\text{-}\tfrac{1}{2} \rangle_e \,. 
\end{flalign}
Specifically, obtaining the measurement $\textbf{S}_e \cdot \hat{n} = -\hbar/2$ projects Eq.~(\ref{eq:transmitted_state}) onto
\begin{flalign}
    |\chi \rangle \equiv& \, _e\langle \Downarrow| \langle j\text{=}N+1| T \rangle \notag \\
    =& \cos(\tfrac{\theta}{2})e^{-i\phi} \sqrt{T_i} |0\rangle - \sin(\tfrac{\theta}{2}) e^{i\phi_+} \sqrt{T_+} |1\rangle \,. \label{eq:logical_state}
\end{flalign}
With some manipulation, we can write Eq.~(\ref{eq:logical_state}) as 
\begin{flalign}
    |\chi \rangle =& p(\tilde{\theta}) \left[ \cos(\tfrac{\tilde{\theta}}{2}) |0 \rangle + \sin(\tfrac{\tilde{\theta}}{2}) e^{i\tilde{\phi}}|1\rangle\, \right] \,
\end{flalign}
where $\tan(\tfrac{\tilde{\theta}}{2}) =\, \sqrt{T_+/T_i}\tan(\tfrac{\theta}{2})$,
$\tilde{\phi} =\,\phi + \phi_+ + \pi $, and
\begin{flalign}
    p(\tilde{\theta}) =& \sqrt{\frac{T_i T_+}{T_+ \cos^2 (\tfrac{\tilde{\theta}}{2})+T_i \sin^2 (\tfrac{\tilde{\theta}}{2})}} \,.
\end{flalign}

\section{SPIN DEPENDENT SCATTERING} 
\label{sec:spindep}

As a simple example of a spin dependent scattering problem, consider the DAE impinging on a scattering region containing a single spin-$s$ particle. When the DAE is in the scattering region, its spin can interact with the spin of the spin-$s$ particle. Ref.~\cite{menezes} treats this problem for the case $s= 1/2$ using the Hamiltonian $H_c = K_{i,c} + \varepsilon_c$. Here the subscript $c$ specifies the continuum case, $K_{i,c}$ is the incoming kinetic energy of the DAE, and the continuum scattering potential is
\begin{equation}
    \varepsilon_c = \frac{J a_c}{\hbar^2} \textbf{S}_e \cdot \textbf{S}_1 \delta (x) \,.
    \label{eq:continuum_potential}
\end{equation}
Note that the interaction strength $J$ has units of energy and $a_c$ is a length scale that will equal the site spacing in the tight-binding case. The incoming kinetic energy of the DAE is given not by the tight-binding dispersion [Eq.~(\ref{eq:dispersion})] but rather
\begin{equation}
    K_{i,c} = \frac{\hbar^2 k_i^2}{2m_e} = t_c k_i^2 a_c^2
    \label{eq:continuum_dispersion}
\end{equation}
where $k_i$ is its the incoming wavenumber, $m_e$ is its mass, and $t_c \equiv \hbar^2/2m_e a_c^2$ is an energy scale that will equal the hopping amplitude in the tight-binding case. Ref.~(\cite{menezes}) finds that the transmission coefficient for the spin flip scattering process $|\downarrow\rangle_e |\uparrow \rangle_1$ $\rightarrow |\uparrow \rangle_e |\downarrow \rangle_1$ is
\begin{flalign}
    T_{f,c} =& \frac{ J_0^2 } {1+\frac{5}{2}J_0^2+ \frac{9}{16}J_0^4 } \,
    \label{eq:continuum_Tf}
\end{flalign}
where $J_0 = \sqrt{2s}J/4t_c k_i a_c$ is unitless.
Likewise, the transmission coefficient for a no spin flip scattering process $|\downarrow\rangle_e |\uparrow \rangle_1$ $\rightarrow |\downarrow \rangle_e |\uparrow\rangle_1$ is 
\begin{flalign}
    T_{nf,c} =& \frac{1+ \frac{1}{4}J_0^2 }{1+\frac{5}{2}J_0^2 + \frac{9}{16}J_0^4 } \label{eq:continuum_Tnf}
\end{flalign}

We now show that the tight-binding Green's function solution we developed in Sec.~\ref{sec:method} replicates the continuum solution, i.e. Eqs.~(\ref{eq:continuum_Tf}) and (\ref{eq:continuum_Tnf}). We then focus on two special cases that were not addressed by the continuum solution, but that our tight-binding solution can handle: inelastic scattering and an interaction with spatial degrees of freedom.

\subsection{Replication of the continuum solution for a contact interaction}
\label{sec:continuum}

\begin{figure}[t]
    \centering
    \includegraphics[width = \linewidth]{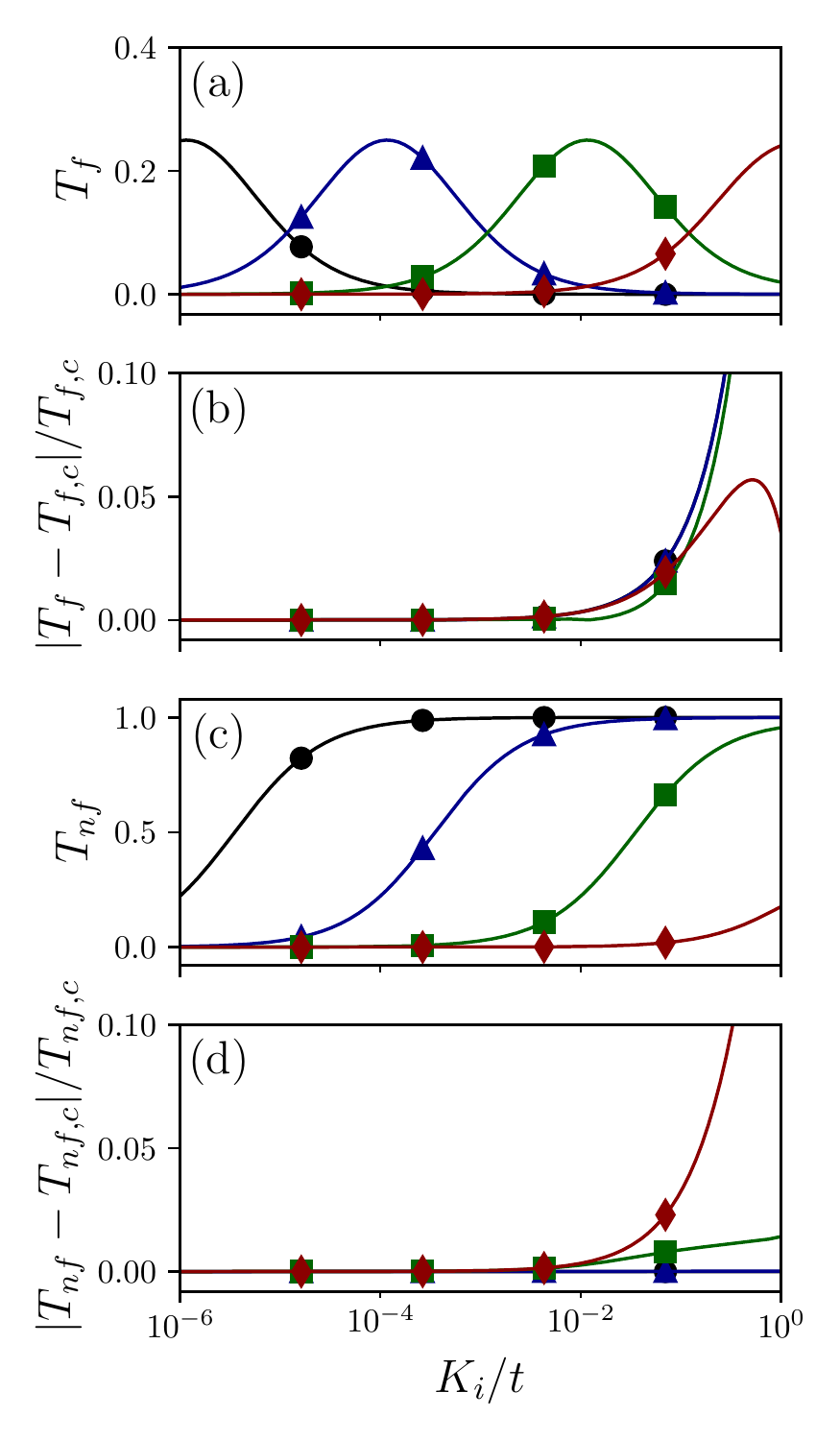}
    \caption{(a) spin flip transmission probability from our tight-binding result and (b) its relative error. (b) no spin flip transmission probability from our tight-binding result and (d) its relative error. Tight-binding parameters are $N=1$ and $t=1.0$ while the interaction strength $J$ is given by (black circles) $J/t= -0.005$, (blue triangles) $J/t=-0.05$, (green squares) $J/t =-0.5$, and (red diamonds) $J/t=-5.0$.}
    \label{fig:continuum}
\end{figure}

We now consider a tight-binding system with site spacing $a=a_c$ and hopping amplitude $t=t_c$. The first task for replicating the continuum solution is to approximate the continuum dispersion, Eq.~(\ref{eq:continuum_dispersion}). For $k_i a \ll 1$, our tight-binding dispersion, Eq.~(\ref{eq:dispersion}), can be written
\begin{flalign}
    K_i =&\, 2t - 2t\left[1 - \frac{1}{2}k_i^2 a^2 + O(k_i^4 a^4) \right] \notag \\
    \approx&\, tk_i^2 a^2 + O(k_i^4 a^4)
\end{flalign}
In other words, our tight-binding dispersion is a good approximation of the continuum dispersion in the case $k_i a \ll 1$. Our baseline expectation is that our results will be a good approximation for $k_i a  \leq 0.1$ corresponding to $K_i / t \leq 0.01$.

The second task for replicating the continuum solution is to approximate the continuum scattering potential, Eq.~(\ref{eq:continuum_potential}). This potential specifies an isotropic exchange interaction, which is a contact interaction in the sense that the DAE only interacts with the spin-1/2 particle when they are at the same point in space. We will approximate this scattering potential by specifying the $\boldsymbol{\varepsilon}_j$ operators $\boldsymbol{\varepsilon}_j$ [which determine the Hamiltonian according to Eq.~(\ref{eq:Hfinite})] as
\begin{equation}
 \boldsymbol{\varepsilon}_j = \frac{J}{\hbar^2} \textbf{S}_e \cdot \textbf{S}_1  \delta_{j1}.
 \label{eq:exchange_single}
\end{equation}
When comparing Eq~(\ref{eq:exchange_single}) to Eq.~(\ref{eq:continuum_potential}), the question to consider is whether the discrete spatial interval $a$ approximates the single continuous point $x=0$. If the DAE's incoming wavelength $2\pi/k_i$ is much larger than the spatial interval $a$, than the DAE should not be sensitive to whether the space is discretized or not. This amounts to $k_i a  \ll 2\pi$ which is covered by the restriction $k_i a \ll 1$ already adopted to replicate the continuum dispersion. We conclude that in the case $k_i a \ll 1$ our tight-binding transmission coefficients, calculated using Eq.~(\ref{eq:Tcoef}), should replicate Eqs.~(\ref{eq:continuum_Tf}) and (\ref{eq:continuum_Tnf}).

To verify this using Eq.~(\ref{eq:exchange_single}), it is convenient to introduce the identity 
$$
\textbf{S}_e \cdot \textbf{S}_1 = S_e^z S_1^z + \frac{1}{2}(S_e^+ S_1^- + S_e^- S_1^+)
$$
where the raising and lowering operators $S_l^\pm$ act on the eigenbasis of a spin-1/2 particle according to $S_l^+ |\uparrow\rangle_l = 0$, $S_l^+ |\downarrow\rangle_l = \hbar |\uparrow\rangle_l$, $S_l^- |\uparrow\rangle_l = \hbar |\downarrow\rangle_l$, and $S_l^- |\downarrow\rangle_l = 0$. As a result, we can write out the action of Eq.~(\ref{eq:exchange_single}) in the two-particle spin space:
\begin{equation}
    \boldsymbol{\varepsilon}_j = \frac{J}{4} \delta_{j1}
    \begin{pmatrix}
    1 & 0 & 0 & 0 \\[2.5pt]
    0 & -1 & 2 & 0  \\[2.5pt]
    0 & 2 & -1& 0  \\[2.5pt]
    0 & 0 & 0 & 1 \\
    \end{pmatrix}
    \begin{matrix}
    |\uparrow \rangle_e |\uparrow \rangle_1\\[2pt]
    |\uparrow\rangle_e |\downarrow \rangle_1\\[2pt]
    |\downarrow\rangle_e |\uparrow \rangle_1\\[2pt]
    |\downarrow\rangle_e |\downarrow \rangle_1\\
    \end{matrix}\,.
    \label{eq:exchange_matrix}
\end{equation}
Using this spin operator, our tight-binding solution yields transmission coefficients that replicate the continuum results when $K_i/t \ll 1$, as shown in Fig.~\ref{fig:continuum}. By looking at the relative error with respect to the continuous results in Figs.~\ref{fig:continuum}(b) and \ref{fig:continuum}(d) we see that the threshold $K_i/t \leq 0.01$ is appropriate. Note that increasing the value of $J$ impacts neither this threshold nor the amplitude of the $T_f$ peaks. However, it does increase both the kinetic energy at which those peaks occur and the $K_i/t$ value at which $T_{nf} \rightarrow 1$.

\subsection{Inelastic scattering}
\label{sec:inelastic_scattering}

\begin{figure}[t]
    \centering
    \includegraphics[width = \linewidth]{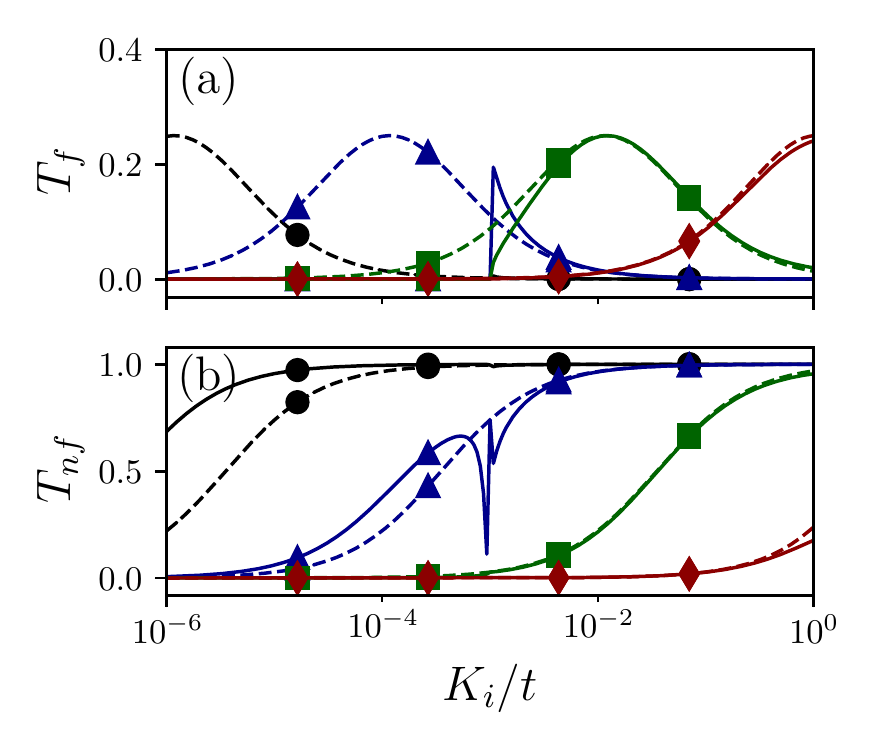}
    \caption{(a) spin flip transmission probability from (solid) our tight-binding result and (dashed) the continuum result $T_{f,c}$. (b) no spin flip transmission probability from (solid) our tight-binding result and (dashed) the continuum result $T_{nf,c}$. Tight-binding parameters are $N=1$, $t=1.0$, and $\Delta/t = 0.001$ while the interaction strength $J$ is given by (black circles) $J/t= -0.005$, (blue triangles) $J/t=-0.05$, (green squares) $J/t =-0.5$, and (red diamonds) $J/t=-5.0$. }
    \label{fig:inelastic}
\end{figure}

Our system of interest involves spin degrees of freedom which may absorb energy during the scattering. As a result, the incoming and outgoing wavenumbers [Eq.~(\ref{eq:wavenumber})] are spin dependent, so inelastic scattering is possible. In this simple example system, inelastic scattering can be accomplished by a Zeeman term on the spin-1/2 particle. This corresponds to adding to Eq.~(\ref{eq:exchange_matrix}) the Zeeman operator
\begin{equation}
    \frac{\Delta}{\hbar} \left( S_1^z+\frac{1}{2}\right) = \Delta     \begin{pmatrix}
    1 & 0 & 0 & 0 \\[2.5pt]
    0 & 0 & 0 & 0  \\[2.5pt]
    0 & 0 & 1 & 0  \\[2.5pt]
    0 & 0 & 0 & 0 \\
    \end{pmatrix}
    \begin{matrix}
    |\uparrow \rangle_e |\uparrow \rangle_1\\[2pt]
    |\uparrow\rangle_e |\downarrow \rangle_1\\[2pt]
    |\downarrow\rangle_e |\uparrow \rangle_1\\[2pt]
    |\downarrow\rangle_e |\downarrow \rangle_1\\
    \end{matrix}\,.
\end{equation}
As shown in Fig.~\ref{fig:inelastic}, the result is that $T_{f}$ is forbidden when $K_i < \Delta$, suddenly turns on at $K_i= \Delta$, and returns to the $\Delta=0$ continuum result when $\Delta \ll K_i \ll t$, before finally starting to diverge from the continuum result again when $K_i \rightarrow t$.

\subsection{Physical origin of the Kondo-like interaction}
\label{sec:physical_origin}

\begin{figure}[t]
    \centering
    \includegraphics[width = \linewidth]{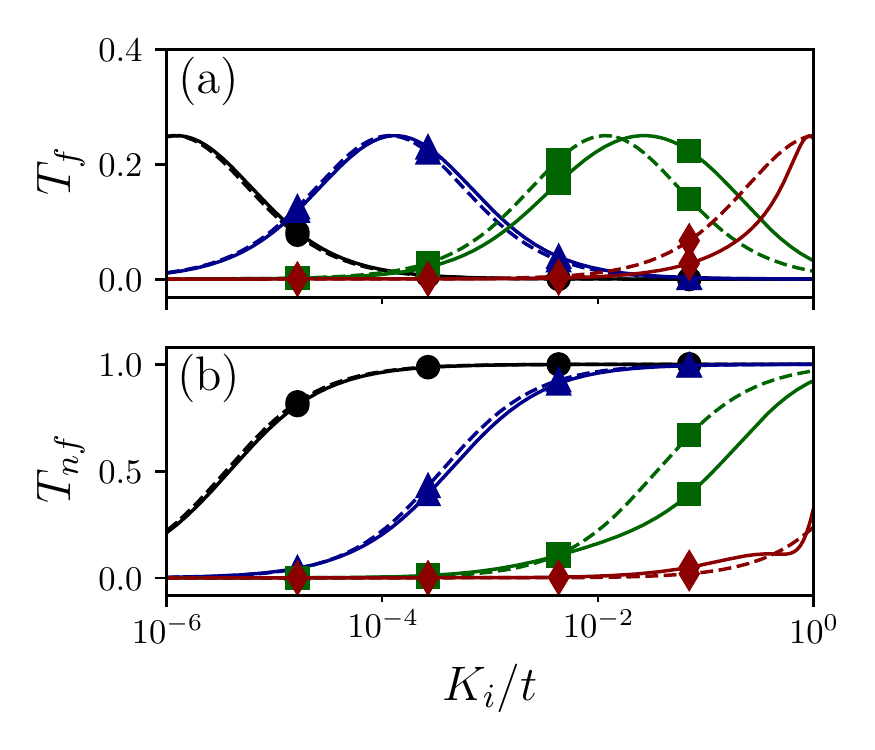}
    \caption{(a) spin flip transmission probability from (solid) our tight-binding result and (dashed) the continuum result $T_{f,c}$. (b) no spin flip transmission probability from (solid) our tight-binding result and (dashed) the continuum result $T_{nf,c}$. Tight-binding parameters are $N=1$, $t=t_h=1.0$, $U_1=0.0$, $U_2 = 100.0$. Using the prefactor of Eq.~(\ref{eq:HSW}) we set (black circles) $\varepsilon=156.2$, $J/t= -0.005$; (blue triangles) $\varepsilon=30.6$, $J/t=-0.05$; (green squares) $\varepsilon = 3.9$, $J/t =-0.5$; and (red diamonds) $\varepsilon = -0.4$, $J/t=-5.0$. }
    \label{fig:origin}
\end{figure}

Our solution specifies the reflection and transmission coefficients [Eqs.~(\ref{eq:Tcoef}) and (\ref{eq:Rcoef})] in terms of the retarded Green's function [Eq.~(\ref{eq:gf_elements})]. As formulated, the $\sigma$ indices encode many-body spin degrees of freedom. However, these indices
could be used to represent many-body quantum numbers
besides spin, so our approach is generally applicable to
interacting Hamiltonians. As an example, consider the Anderson model, which describes the Coulomb interaction between a conduction band electron and an electron in a localized orbital~\cite{schrieffer}. The Anderson model describes the physical origin of the Kondo-like interaction we have used throughout because a Schreiffer-Wolff transformation of this model allows the electrons to interact via their spins rather than their charges~\cite{schrieffer}. 

Consider a scattering region of size $N=2$ containing two sites $j=1,2$ with hopping $t_h$ between them. A single electron occupying site $j$ has energy $\varepsilon_j$ and two electrons occupying site $j$ experience Coulomb repulsion $U_j$. For a system of two antiparallel electrons, the Anderson Hamiltonian is \cite{koch}
\begin{equation}
    H_A =  \varepsilon \hat{I}+    \begin{pmatrix}
    U_1-\varepsilon  & -t_h & t_h  & 0 \\
    -t_h  & 0 & 0  & -t_h \\
    t_h  & 0 & 0  & t_h  \\
    0  & -t_h & t_h  & U_2+\varepsilon \\
    \end{pmatrix}
    \begin{matrix}
    |\uparrow\downarrow\rangle_1 |\,\rangle_2 \\ |\uparrow\rangle_1 |\downarrow\rangle_2 \\ |\downarrow\rangle_1 |\uparrow\rangle_2 \\ |\,\rangle_1 |\uparrow \downarrow \rangle_2
    \end{matrix}\,,
    \label{eq:Handerson}
\end{equation}
where $\varepsilon \equiv \varepsilon_2 - \varepsilon_1$. Care must be taken before inserting Eq.~(\ref{eq:Handerson}) directly into the Hamiltonian [Eq.~(\ref{eq:Hfinite})] because the `spin' basis
$|\sigma \rangle = \{ |\uparrow\downarrow\rangle_1 |\rangle_2$,
$|\uparrow\rangle_1 |\downarrow\rangle_2$,
$|\downarrow\rangle_2 |\uparrow\rangle_2 $,
$|\rangle_1 |\uparrow \downarrow \rangle_2  \}$ also contains spatial degrees of freedom in the doubly occupied states. We plainly cannot `transmit' the state $|\uparrow \downarrow \rangle_1 |\rangle_2$.

The first way around this is to interpret the system as a one dimensional chain of sites $j=...0,1,3...$ where the site $j=1$ is coupled to an off-chain site $j=2$ which together form the scattering region. Both electrons can move freely in the scattering region, but only one electron can continue into the leads. Mathematically, this corresponds to modifying the hopping matrix $\textbf{t}=t\textbf{I}$ in Eq.~(\ref{eq:Hfinite}) to $\textbf{t} = \text{diag}(0,t,t,0)$. 

Alternatively, we can ignore the double occupancy states, which are higher in energy, instead focusing on the lower energy subspace $\{ |\uparrow\rangle_1 |\downarrow\rangle_2$,
$|\downarrow\rangle_2 |\uparrow\rangle_2 \}$. Mathematically, this is accomplished by a Schrieffer-Wolff transformation~\cite{schrieffer}, an expansion to second order in the small quantities  $t_h/|U_1 - \varepsilon |$ and $t_h/|U_2 + \varepsilon |$ \cite{koch}. The resulting Schrieffer-Wolff Hamiltonian is \cite{koch}
\begin{flalign}
    H_{SW} =& \left[ \frac{-t_h^2 (U_1 + U_2)}{(U_1 - \varepsilon)(U_2 + \varepsilon)}\right]
\begin{pmatrix}
1 & -1  \\
-1 & 1 \\
\end{pmatrix}\begin{matrix}
|\uparrow \rangle_1 |\downarrow \rangle_2 \\ |\downarrow \rangle_1 |\uparrow \rangle_2 \end{matrix} \notag \\
=&  \left[ \frac{2t_h^2 (U_1 + U_2)}{(U_1 - \varepsilon)(U_2 + \varepsilon)}\right] \left( \frac{1}{\hbar^2} \textbf{S}_1 \cdot \textbf{S}_2 - \frac{1}{4} \textbf{I} \right)\,.
\label{eq:HSW}
\end{flalign}
This is an exchange interaction with the same dot product of spin operators we have used for the Kondo-like interaction throughout this paper. The quantity in square brackets defines the interaction strength $J$. Note that for $U_1 = U_2 = 0$, $J=0$, revealing that the interaction is rooted in the Coulomb repulsion between the two electrons. Contrary to the geometrical interpretation of Eq.~(\ref{eq:Handerson}), in Eq.~(\ref{eq:HSW}) the spatial degrees of freedom of the off-chain site $j=2$ have been combined with the spin degrees of freedom of the on-chain site $j=1$. In other words, the system has an effectively one-dimensional geometry.

Our solution allows us to implement either the exact Anderson Hamiltonian by setting $\boldsymbol{\varepsilon}_j =(H_A + \tfrac{J}{4}\textbf{I}) \delta_{1j}$
or the perturbative Schreiffer-Wolff Hamiltonian by setting $\boldsymbol{\varepsilon}_j = (H_{SW}+\tfrac{J}{4}\textbf{I}) \delta_{1j}$, removing the constant energy shift in both cases to isolate the dot product of spin operators. We took the latter approach throughout this paper and compared it to the continuum results it in Fig.~\ref{fig:continuum}. In Fig.~\ref{fig:origin} we compare the former approach to the continuum results at different values of $J/t$. We specify $J/t$ by setting $\varepsilon$, and this determines the small quantities $t_h/|U_1-\varepsilon|$ and $t_h/|U_2 + \varepsilon|$. One of these quantities always increases with increasing $J$ so that the agreement is poorer. Indeed, in Fig.~\ref{fig:origin} we see that the breakdown in the agreement at larger $K_i/t$ is made worse when $J/t$ is larger. 

\bibliography{aps}

\begin{thebibliography}{38}%
\makeatletter
\providecommand \@ifxundefined [1]{%
 \@ifx{#1\undefined}
}%
\providecommand \@ifnum [1]{%
 \ifnum #1\expandafter \@firstoftwo
 \else \expandafter \@secondoftwo
 \fi
}%
\providecommand \@ifx [1]{%
 \ifx #1\expandafter \@firstoftwo
 \else \expandafter \@secondoftwo
 \fi
}%
\providecommand \natexlab [1]{#1}%
\providecommand \enquote  [1]{``#1''}%
\providecommand \bibnamefont  [1]{#1}%
\providecommand \bibfnamefont [1]{#1}%
\providecommand \citenamefont [1]{#1}%
\providecommand \href@noop [0]{\@secondoftwo}%
\providecommand \href [0]{\begingroup \@sanitize@url \@href}%
\providecommand \@href[1]{\@@startlink{#1}\@@href}%
\providecommand \@@href[1]{\endgroup#1\@@endlink}%
\providecommand \@sanitize@url [0]{\catcode `\\12\catcode `\$12\catcode
  `\&12\catcode `\#12\catcode `\^12\catcode `\_12\catcode `\%12\relax}%
\providecommand \@@startlink[1]{}%
\providecommand \@@endlink[0]{}%
\providecommand \url  [0]{\begingroup\@sanitize@url \@url }%
\providecommand \@url [1]{\endgroup\@href {#1}{\urlprefix }}%
\providecommand \urlprefix  [0]{URL }%
\providecommand \Eprint [0]{\href }%
\providecommand \doibase [0]{https://doi.org/}%
\providecommand \selectlanguage [0]{\@gobble}%
\providecommand \bibinfo  [0]{\@secondoftwo}%
\providecommand \bibfield  [0]{\@secondoftwo}%
\providecommand \translation [1]{[#1]}%
\providecommand \BibitemOpen [0]{}%
\providecommand \bibitemStop [0]{}%
\providecommand \bibitemNoStop [0]{.\EOS\space}%
\providecommand \EOS [0]{\spacefactor3000\relax}%
\providecommand \BibitemShut  [1]{\csname bibitem#1\endcsname}%
\let\auto@bib@innerbib\@empty
\bibitem [{\citenamefont {{Michael Nielsen and Isaac Chuang}}(2011)}]{nielsen}%
  \BibitemOpen
  \bibfield  {author} {\bibinfo {author} {\bibnamefont {{Michael Nielsen and
  Isaac Chuang}}},\ }\href@noop {} {\emph {\bibinfo {title} {Quantum
  Computation and Quantum Information: 10th Anniversary Edition}}}\ (\bibinfo
  {publisher} {Cambridge University Press},\ \bibinfo {address} {Cambridge},\
  \bibinfo {year} {2011})\BibitemShut {NoStop}%
\bibitem [{\citenamefont {{Charles H. Bennett and David P.
  DiVincenzo}}(2000)}]{div_qis}%
  \BibitemOpen
  \bibfield  {author} {\bibinfo {author} {\bibnamefont {{Charles H. Bennett and
  David P. DiVincenzo}}},\ }\bibfield  {title} {\bibinfo {title}
  {\href{https://www.nature.com/articles/35005001}{Quantum information and
  computation}},\ }\href@noop {} {\bibfield  {journal} {\bibinfo  {journal}
  {Nature}\ }\textbf {\bibinfo {volume} {404}},\ \bibinfo {pages} {247}
  (\bibinfo {year} {2000})}\BibitemShut {NoStop}%
\bibitem [{\citenamefont {{Giuseppe Castagnoli and David Ritz
  Finkelstein}}(2001)}]{castagnoli}%
  \BibitemOpen
  \bibfield  {author} {\bibinfo {author} {\bibnamefont {{Giuseppe Castagnoli
  and David Ritz Finkelstein}}},\ }\bibfield  {title} {\bibinfo {title}
  {\href{https://doi.org/10.1098/rspa.2001.0797}{Theory of the quantum
  speed-up}},\ }\href@noop {} {\bibfield  {journal} {\bibinfo  {journal} {Proc.
  R. Soc. A}\ }\textbf {\bibinfo {volume} {457}},\ \bibinfo {pages} {1799}
  (\bibinfo {year} {2001})}\BibitemShut {NoStop}%
\bibitem [{\citenamefont {{Matteo Atzori and Roberta
  Sessoli}}(2019)}]{sessoli}%
  \BibitemOpen
  \bibfield  {author} {\bibinfo {author} {\bibnamefont {{Matteo Atzori and
  Roberta Sessoli}}},\ }\bibfield  {title} {\bibinfo {title}
  {\href{https://doi.org/10.1021/jacs.9b00984}{The Second Quantum Revolution:
  Role and Challenges of Molecular Chemistry}},\ }\href@noop {} {\bibfield
  {journal} {\bibinfo  {journal} {J. Am. Chem. Soc.}\ }\textbf {\bibinfo
  {volume} {141}},\ \bibinfo {pages} {11339} (\bibinfo {year}
  {2019})}\BibitemShut {NoStop}%
\bibitem [{\citenamefont {{Matteo Atzori, Elena Morra, Lorenzo Tesi, Andrea
  Albino, Mario Chiesa, Lorenzo Sorace, and Roberta
  Sessoli}}(2016)}]{sessoli_van}%
  \BibitemOpen
  \bibfield  {author} {\bibinfo {author} {\bibnamefont {{Matteo Atzori, Elena
  Morra, Lorenzo Tesi, Andrea Albino, Mario Chiesa, Lorenzo Sorace, and Roberta
  Sessoli}}},\ }\bibfield  {title} {\bibinfo {title}
  {\href{https://doi.org/10.1021/jacs.6b05574}{Quantum Coherence Times
  Enhancement in Vanadium(IV)-based Potential Molecular Qubits: the Key Role of
  the Vanadyl Moiety}},\ }\href@noop {} {\bibfield  {journal} {\bibinfo
  {journal} {J. Am. Chem. Soc.}\ }\textbf {\bibinfo {volume} {138}},\ \bibinfo
  {pages} {11234–11244} (\bibinfo {year} {2016})}\BibitemShut {NoStop}%
\bibitem [{\citenamefont {{Joseph M. Zadrozny, Jens Niklas, Oleg G. Poluektov,
  and Danna E. Freedman}}(2015)}]{zadrozny}%
  \BibitemOpen
  \bibfield  {author} {\bibinfo {author} {\bibnamefont {{Joseph M. Zadrozny,
  Jens Niklas, Oleg G. Poluektov, and Danna E. Freedman}}},\ }\bibfield
  {title} {\bibinfo {title}
  {\href{https://doi.org/10.1021/acscentsci.5b00338}{Millisecond Coherence Time
  in a Tunable Molecular Electronic Spin Qubit}},\ }\href@noop {} {\bibfield
  {journal} {\bibinfo  {journal} {ACS Cent. Sci.}\ }\textbf {\bibinfo {volume}
  {1}},\ \bibinfo {pages} {488} (\bibinfo {year} {2015})}\BibitemShut {NoStop}%
\bibitem [{\citenamefont {{A. Gaita-Ariño, F. Luis, S. Hill, E.
  Coronado}}(2019)}]{hill}%
  \BibitemOpen
  \bibfield  {author} {\bibinfo {author} {\bibnamefont {{A. Gaita-Ariño, F.
  Luis, S. Hill, E. Coronado}}},\ }\bibfield  {title} {\bibinfo {title}
  {\href{https://doi.org/10.1038/s41557-019-0232-y}{Molecular spins for quantum
  computation}},\ }\href@noop {} {\bibfield  {journal} {\bibinfo  {journal}
  {Nat. Chem.}\ }\textbf {\bibinfo {volume} {11}},\ \bibinfo {pages} {301}
  (\bibinfo {year} {2019})}\BibitemShut {NoStop}%
\bibitem [{\citenamefont {{Daniel Loss and David P.
  DiVincenzo}}(1998)}]{div_loss}%
  \BibitemOpen
  \bibfield  {author} {\bibinfo {author} {\bibnamefont {{Daniel Loss and David
  P. DiVincenzo}}},\ }\bibfield  {title} {\bibinfo {title}
  {\href{https://doi.org/10.1103/PhysRevA.57.120}{Quantum computation with
  quantum dots}},\ }\href@noop {} {\bibfield  {journal} {\bibinfo  {journal}
  {Phys. Rev. A}\ }\textbf {\bibinfo {volume} {57}},\ \bibinfo {pages} {120}
  (\bibinfo {year} {1998})}\BibitemShut {NoStop}%
\bibitem [{\citenamefont {{A. T. Costa, Jr., S. Bose, Y. Omar}}(2006)}]{costa}%
  \BibitemOpen
  \bibfield  {author} {\bibinfo {author} {\bibnamefont {{A. T. Costa, Jr., S.
  Bose, Y. Omar}}},\ }\bibfield  {title} {\bibinfo {title}
  {\href{https://journals.aps.org/prl/abstract/10.1103/PhysRevLett.96.230501}{Entanglement
  of Two Impurities through Electron Scattering}},\ }\href@noop {} {\bibfield
  {journal} {\bibinfo  {journal} {Phys. Rev. Lett.}\ }\textbf {\bibinfo
  {volume} {96}},\ \bibinfo {pages} {230501} (\bibinfo {year}
  {2006})}\BibitemShut {NoStop}%
\bibitem [{\citenamefont {{ Eric D. Switzer, Xiao-Guang Zhang, Talat S.
  Rahman}}(2021)}]{switzer}%
  \BibitemOpen
  \bibfield  {author} {\bibinfo {author} {\bibnamefont {{ Eric D. Switzer,
  Xiao-Guang Zhang, Talat S. Rahman}}},\ }\bibfield  {title} {\bibinfo {title}
  {\href{https://doi.org/10.1103/PhysRevA.104.052434}{Anisotropy-Exchange
  Resonance as a Mechanism for Entangled State Switching}},\ }\href@noop {}
  {\bibfield  {journal} {\bibinfo  {journal} {Phys. Rev. A}\ }\textbf {\bibinfo
  {volume} {104}},\ \bibinfo {pages} {052434} (\bibinfo {year}
  {2021})}\BibitemShut {NoStop}%
\bibitem [{\citenamefont {{ Eric D. Switzer, Xiao-Guang Zhang, Talat S.
  Rahman}}(2022)}]{switzer2}%
  \BibitemOpen
  \bibfield  {author} {\bibinfo {author} {\bibnamefont {{ Eric D. Switzer,
  Xiao-Guang Zhang, Talat S. Rahman}}},\ }\bibfield  {title} {\bibinfo {title}
  {\href{https://doi.org/10.1088/2399-6528/ac7e1d}{Electronic control and
  switching of entangled spin state using anisotropy and exchange in the
  three-particle paradigm}},\ }\href@noop {} {\bibfield  {journal} {\bibinfo
  {journal} {J. Phys. Commun.}\ }\textbf {\bibinfo {volume} {6}},\ \bibinfo
  {pages} {075007} (\bibinfo {year} {2022})}\BibitemShut {NoStop}%
\bibitem [{\citenamefont {{K. Yuasa, H. Nakazato}}(2006)}]{yuasa}%
  \BibitemOpen
  \bibfield  {author} {\bibinfo {author} {\bibnamefont {{K. Yuasa, H.
  Nakazato}}},\ }\bibfield  {title} {\bibinfo {title}
  {\href{https://iopscience.iop.org/article/10.1088/1751-8113/40/2/009}{Resonant
  scattering can enhance the degree of entanglement}},\ }\href@noop {}
  {\bibfield  {journal} {\bibinfo  {journal} {J. Phys. A: Math. Theor.}\
  }\textbf {\bibinfo {volume} {40}},\ \bibinfo {pages} {297} (\bibinfo {year}
  {2006})}\BibitemShut {NoStop}%
\bibitem [{\citenamefont {{Francesco Ciccarello, Massimo Palma, Michelangelo
  Zarcone, Yasser Omar, Vitor Rocha Vieira}}(2006)}]{ciccarello}%
  \BibitemOpen
  \bibfield  {author} {\bibinfo {author} {\bibnamefont {{Francesco Ciccarello,
  Massimo Palma, Michelangelo Zarcone, Yasser Omar, Vitor Rocha Vieira}}},\
  }\bibfield  {title} {\bibinfo {title}
  {\href{https://iopscience.iop.org/article/10.1088/1367-2630/8/9/214}{Entanglement
  controlled single-electron transmittivity}},\ }\href@noop {} {\bibfield
  {journal} {\bibinfo  {journal} {New J. Phys.}\ }\textbf {\bibinfo {volume}
  {8}},\ \bibinfo {pages} {214} (\bibinfo {year} {2006})}\BibitemShut {NoStop}%
\bibitem [{\citenamefont {{Francesco Ciccarello, G. Massimo Palma, Mauro
  Paternostro, Michelangelo Zarcone, Yasser Omar}}(2009)}]{ciccarello2}%
  \BibitemOpen
  \bibfield  {author} {\bibinfo {author} {\bibnamefont {{Francesco Ciccarello,
  G. Massimo Palma, Mauro Paternostro, Michelangelo Zarcone, Yasser Omar}}},\
  }\bibfield  {title} {\bibinfo {title}
  {\href{https://doi.org/10.1016/J.SOLIDSTATESCIENCES.2008.01.012}{Entanglement
  generation between two spin-s magnetic impurities in a solid via electron
  scattering}},\ }\href@noop {} {\bibfield  {journal} {\bibinfo  {journal}
  {Solid State Sciences}\ }\textbf {\bibinfo {volume} {11}},\ \bibinfo {pages}
  {931} (\bibinfo {year} {2009})}\BibitemShut {NoStop}%
\bibitem [{\citenamefont {{G. Cordourier-Maruri, F. Ciccarello, Y. Omar, M.
  Zarcone, R. de Coss, S. Bose}}(2010)}]{maruri}%
  \BibitemOpen
  \bibfield  {author} {\bibinfo {author} {\bibnamefont {{G. Cordourier-Maruri,
  F. Ciccarello, Y. Omar, M. Zarcone, R. de Coss, S. Bose}}},\ }\bibfield
  {title} {\bibinfo {title}
  {\href{http://dx.doi.org/10.1103/PhysRevA.82.052313}{Implementing quantum
  gates through scattering between a static and a flying qubit}},\ }\href@noop
  {} {\bibfield  {journal} {\bibinfo  {journal} {Phys. Rev. A}\ }\textbf
  {\bibinfo {volume} {82}},\ \bibinfo {pages} {052313} (\bibinfo {year}
  {2010})}\BibitemShut {NoStop}%
\bibitem [{\citenamefont {{F. Ciccarello, D. E. Browne, L. C. Kwek, H.
  Schomerus, M. Zarcone, S. Bose}}(2012)}]{ciccarello_gate}%
  \BibitemOpen
  \bibfield  {author} {\bibinfo {author} {\bibnamefont {{F. Ciccarello, D. E.
  Browne, L. C. Kwek, H. Schomerus, M. Zarcone, S. Bose}}},\ }\bibfield
  {title} {\bibinfo {title}
  {\href{https://doi.org/10.1103/PhysRevA.85.050305}{Quasideterministic
  realization of a universal quantum gate in a single scattering process}},\
  }\href@noop {} {\bibfield  {journal} {\bibinfo  {journal} {Phys. Rev. A}\
  }\textbf {\bibinfo {volume} {85}},\ \bibinfo {pages} {050305} (\bibinfo
  {year} {2012})}\BibitemShut {NoStop}%
\bibitem [{\citenamefont {{ O. L. T. de Menezes, J. S.
  Helman}}(1985)}]{menezes}%
  \BibitemOpen
  \bibfield  {author} {\bibinfo {author} {\bibnamefont {{ O. L. T. de Menezes,
  J. S. Helman}}},\ }\bibfield  {title} {\bibinfo {title}
  {\href{https://doi.org/10.1119/1.14041}{Spin flip enhancement at resonant
  transmission}},\ }\href@noop {} {\bibfield  {journal} {\bibinfo  {journal}
  {Am. J. Phys.}\ }\textbf {\bibinfo {volume} {53}},\ \bibinfo {pages} {1100}
  (\bibinfo {year} {1985})}\BibitemShut {NoStop}%
\bibitem [{\citenamefont {{{P. A. Khomyakov, G. Brocks, V. Karpan, M.
  Zwierzycki, P. J. Kelly}}}(2005)}]{khomyakov}%
  \BibitemOpen
  \bibfield  {author} {\bibinfo {author} {\bibnamefont {{{P. A. Khomyakov, G.
  Brocks, V. Karpan, M. Zwierzycki, P. J. Kelly}}}},\ }\bibfield  {title}
  {\bibinfo {title}
  {\href{http://dx.doi.org/10.1103/PhysRevB.72.035450}{Conductance calculations
  for quantum wires and interfaces: Mode matching and Green’s functions}},\
  }\href@noop {} {\bibfield  {journal} {\bibinfo  {journal} {Phys. Rev. B}\
  }\textbf {\bibinfo {volume} {72}},\ \bibinfo {pages} {035450} (\bibinfo
  {year} {2005})}\BibitemShut {NoStop}%
\bibitem [{\citenamefont {{Sander J. Tans, Michel H. Devoret, Hongjie Dai,
  Andreas Thess, Richard E. Smalley, L. J. Geerligs and Cees
  Dekker}}(1997)}]{tans}%
  \BibitemOpen
  \bibfield  {author} {\bibinfo {author} {\bibnamefont {{Sander J. Tans, Michel
  H. Devoret, Hongjie Dai, Andreas Thess, Richard E. Smalley, L. J. Geerligs
  and Cees Dekker}}},\ }\bibfield  {title} {\bibinfo {title}
  {\href{https://www.nature.com/articles/386474a0}{Individual single-wall
  carbon nanotubes as quantum wires}},\ }\href@noop {} {\bibfield  {journal}
  {\bibinfo  {journal} {Nature}\ }\textbf {\bibinfo {volume} {386}},\ \bibinfo
  {pages} {474} (\bibinfo {year} {1997})}\BibitemShut {NoStop}%
\bibitem [{\citenamefont {{M.-V. Fernandez-Serra, Ch. Adessi, and X.
  Blase}}(2006)}]{serra}%
  \BibitemOpen
  \bibfield  {author} {\bibinfo {author} {\bibnamefont {{M.-V. Fernandez-Serra,
  Ch. Adessi, and X. Blase}}},\ }\bibfield  {title} {\bibinfo {title}
  {\href{https://doi.org/10.1021/nl0614258}{Conductance, Surface Traps, and
  Passivation in Doped Silicon Nanowires}},\ }\href@noop {} {\bibfield
  {journal} {\bibinfo  {journal} {Nano Lett.}\ }\textbf {\bibinfo {volume}
  {6}},\ \bibinfo {pages} {2674} (\bibinfo {year} {2006})}\BibitemShut
  {NoStop}%
\bibitem [{\citenamefont {{M. Urdampilleta, S. Klyatskaya, J-P. Cleuziou, M.
  Ruben, and W. Wernsdorfer}}(2011)}]{wernsdorfer}%
  \BibitemOpen
  \bibfield  {author} {\bibinfo {author} {\bibnamefont {{M. Urdampilleta, S.
  Klyatskaya, J-P. Cleuziou, M. Ruben, and W. Wernsdorfer}}},\ }\bibfield
  {title} {\bibinfo {title}
  {\href{https://www.nature.com/articles/nmat3050}{Supramolecular spin
  valves}},\ }\href@noop {} {\bibfield  {journal} {\bibinfo  {journal} {Nature
  Mater.}\ }\textbf {\bibinfo {volume} {10}},\ \bibinfo {pages} {502} (\bibinfo
  {year} {2011})}\BibitemShut {NoStop}%
\bibitem [{\citenamefont {{H. Aurich, A. Baumgartner, F. Freitag, A. Eichler,
  J. Trbovic, and C. Schönenberger}}(2010)}]{baumgartner}%
  \BibitemOpen
  \bibfield  {author} {\bibinfo {author} {\bibnamefont {{H. Aurich, A.
  Baumgartner, F. Freitag, A. Eichler, J. Trbovic, and C. Schönenberger}}},\
  }\bibfield  {title} {\bibinfo {title}
  {\href{https://doi.org/10.1063/1.3502600}{Permalloy-based carbon nanotube
  spin-valve}},\ }\href@noop {} {\bibfield  {journal} {\bibinfo  {journal}
  {Appl. Phys. Lett.}\ }\textbf {\bibinfo {volume} {97}},\ \bibinfo {pages}
  {153116} (\bibinfo {year} {2010})}\BibitemShut {NoStop}%
\bibitem [{\citenamefont {{David Kalkstein and Paul Soven}}(1971)}]{soven}%
  \BibitemOpen
  \bibfield  {author} {\bibinfo {author} {\bibnamefont {{David Kalkstein and
  Paul Soven}}},\ }\bibfield  {title} {\bibinfo {title}
  {\href{https://www.sciencedirect.com/science/article/pii/0039602871901154}{A
  green's function theory of surface states}},\ }\href@noop {} {\bibfield
  {journal} {\bibinfo  {journal} {Surf. Sci.}\ }\textbf {\bibinfo {volume}
  {26}},\ \bibinfo {pages} {85} (\bibinfo {year} {1971})}\BibitemShut {NoStop}%
\bibitem [{\citenamefont {{Julian Velev and William Butler}}(2004)}]{butler}%
  \BibitemOpen
  \bibfield  {author} {\bibinfo {author} {\bibnamefont {{Julian Velev and
  William Butler}}},\ }\bibfield  {title} {\bibinfo {title}
  {\href{https://iopscience.iop.org/article/10.1088/0953-8984/16/21/R01}{On the
  equivalence of different techniques for evaluating the Green function for a
  semi-infinite system using a localized basis}},\ }\href@noop {} {\bibfield
  {journal} {\bibinfo  {journal} {J. Phys. Condens. Matter}\ }\textbf {\bibinfo
  {volume} {16}},\ \bibinfo {pages} {R637–R657} (\bibinfo {year}
  {2004})}\BibitemShut {NoStop}%
\bibitem [{\citenamefont {{R. Haydock, V. Heine, M. Kelly}}(1972)}]{haydock}%
  \BibitemOpen
  \bibfield  {author} {\bibinfo {author} {\bibnamefont {{R. Haydock, V. Heine,
  M. Kelly}}},\ }\bibfield  {title} {\bibinfo {title}
  {\href{https://iopscience.iop.org/article/10.1088/0022-3719/5/20/004}{Electronic
  structure based on the local atomic environment for tight-binding bands}},\
  }\href@noop {} {\bibfield  {journal} {\bibinfo  {journal} {J. Phys. C: Solid
  State Phys.}\ }\textbf {\bibinfo {volume} {5}},\ \bibinfo {pages} {2845}
  (\bibinfo {year} {1972})}\BibitemShut {NoStop}%
\bibitem [{\citenamefont {{A. MacKinnon}}(1985)}]{mackinnon}%
  \BibitemOpen
  \bibfield  {author} {\bibinfo {author} {\bibnamefont {{A. MacKinnon}}},\
  }\bibfield  {title} {\bibinfo {title}
  {\href{https://link.springer.com/content/pdf/10.1007/BF01328846.pdf}{The
  Calculation of Transport Properties and Density of States of Disordered
  Solids}},\ }\href@noop {} {\bibfield  {journal} {\bibinfo  {journal} {Z.
  Phys. B}\ }\textbf {\bibinfo {volume} {59}},\ \bibinfo {pages} {385}
  (\bibinfo {year} {1985})}\BibitemShut {NoStop}%
\bibitem [{\citenamefont {{J. R. Schrieffer and P. A.
  Wolff}}(1966)}]{schrieffer}%
  \BibitemOpen
  \bibfield  {author} {\bibinfo {author} {\bibnamefont {{J. R. Schrieffer and
  P. A. Wolff}}},\ }\bibfield  {title} {\bibinfo {title}
  {\href{https://doi.org/10.1103/PhysRev.149.491}{Relation between the Anderson
  and Kondo Hamiltonians}},\ }\href@noop {} {\bibfield  {journal} {\bibinfo
  {journal} {Phys. Rev.}\ }\textbf {\bibinfo {volume} {149}},\ \bibinfo {pages}
  {491} (\bibinfo {year} {1966})}\BibitemShut {NoStop}%
\bibitem [{\citenamefont {Koch}(2017)}]{koch}%
  \BibitemOpen
  \bibfield  {author} {\bibinfo {author} {\bibfnamefont {E.}~\bibnamefont
  {Koch}},\ }\bibfield  {title} {\bibinfo {title}
  {\href{https://www.cond-mat.de/events/correl17/manuscripts/koch.pdf}{The
  Physics of Correlated Insulators, Metals, and Superconductors}}\ }(\bibinfo
  {publisher} {Verlag des Forschungszentrum Jülich},\ \bibinfo {address}
  {Jülich},\ \bibinfo {year} {2017})\ Chap.~\bibinfo {chapter} {4}\BibitemShut
  {NoStop}%
\bibitem [{\citenamefont {{Romain Vincent, Svetlana Klyatskaya, Mario Ruben,
  Wolfgang Wernsdorfer \& Franck Balestro}}(2012)}]{wernsdorfer_nuc}%
  \BibitemOpen
  \bibfield  {author} {\bibinfo {author} {\bibnamefont {{Romain Vincent,
  Svetlana Klyatskaya, Mario Ruben, Wolfgang Wernsdorfer \& Franck
  Balestro}}},\ }\bibfield  {title} {\bibinfo {title}
  {\href{https://www.nature.com/articles/nature11341}{Electronic read-out of a
  single nuclear spin using a molecular spin transistor}},\ }\href@noop {}
  {\bibfield  {journal} {\bibinfo  {journal} {Nature}\ }\textbf {\bibinfo
  {volume} {488}},\ \bibinfo {pages} {357} (\bibinfo {year}
  {2012})}\BibitemShut {NoStop}%
\bibitem [{\citenamefont {{Jie-Xiang Yu, George Christou, and Hai-Ping
  Cheng}}(2020)}]{yu}%
  \BibitemOpen
  \bibfield  {author} {\bibinfo {author} {\bibnamefont {{Jie-Xiang Yu, George
  Christou, and Hai-Ping Cheng}}},\ }\bibfield  {title} {\bibinfo {title}
  {\href{https://dx.doi.org/10.1021/acs.jpcc.0c02213}{Analysis of Exchange
  Interactions in Dimers of Mn3 Single-Molecule Magnets, and Their Sensitivity
  to External Pressure}},\ }\href@noop {} {\bibfield  {journal} {\bibinfo
  {journal} {J. Phys. Chem. C}\ }\textbf {\bibinfo {volume} {124}},\ \bibinfo
  {pages} {14768} (\bibinfo {year} {2020})}\BibitemShut {NoStop}%
\bibitem [{\citenamefont {{Haechan Park, Shuanglong Liu, James N. Fry, and
  Hai-Ping Cheng}}(2022)}]{park}%
  \BibitemOpen
  \bibfield  {author} {\bibinfo {author} {\bibnamefont {{Haechan Park,
  Shuanglong Liu, James N. Fry, and Hai-Ping Cheng}}},\ }\bibfield  {title}
  {\bibinfo {title}
  {\href{https://doi.org/10.1103/PhysRevB.105.195408}{First-principles study of
  bilayer polymeric manganese phthalocyanine}},\ }\href@noop {} {\bibfield
  {journal} {\bibinfo  {journal} {Phys. Rev. B}\ }\textbf {\bibinfo {volume}
  {105}},\ \bibinfo {pages} {195408} (\bibinfo {year} {2022})}\BibitemShut
  {NoStop}%
\bibitem [{\citenamefont {{Gopalan Rajaraman, E. Carolina Sañudo, Madeleine
  Helliwell, Stergios Piligkos, Wolfgang Wernsdorfer, George Christou, Euan
  K.Brechin}}(2005)}]{christou_MnIII}%
  \BibitemOpen
  \bibfield  {author} {\bibinfo {author} {\bibnamefont {{Gopalan Rajaraman, E.
  Carolina Sañudo, Madeleine Helliwell, Stergios Piligkos, Wolfgang
  Wernsdorfer, George Christou, Euan K.Brechin}}},\ }\bibfield  {title}
  {\bibinfo {title} {\href{https://doi.org/10.1016/j.poly.2005.03.046}{Magnetic
  and theoretical characterization of a ferromagnetic Mn(III) dimer}},\
  }\href@noop {} {\bibfield  {journal} {\bibinfo  {journal} {Polyhedron}\
  }\textbf {\bibinfo {volume} {24}},\ \bibinfo {pages} {2450} (\bibinfo {year}
  {2005})}\BibitemShut {NoStop}%
\bibitem [{\citenamefont {{Wolfgang Wernsdorfer, Núria Aliaga-Alcalde, David
  N. Hendrickson \& George Christou}}(2002)}]{christou_Mn4}%
  \BibitemOpen
  \bibfield  {author} {\bibinfo {author} {\bibnamefont {{Wolfgang Wernsdorfer,
  Núria Aliaga-Alcalde, David N. Hendrickson \& George Christou}}},\
  }\bibfield  {title} {\bibinfo {title}
  {\href{https://www.nature.com/articles/416406a}{Exchange-biased quantum
  tunnelling in a supramolecular dimer of single-molecule magnets}},\
  }\href@noop {} {\bibfield  {journal} {\bibinfo  {journal} {Nature}\ }\textbf
  {\bibinfo {volume} {416}},\ \bibinfo {pages} {406} (\bibinfo {year}
  {2002})}\BibitemShut {NoStop}%
\bibitem [{\citenamefont {{Zahra Hooshmand, Jie-Xiang Yu, Hai-Ping Cheng, and
  Mark R Pederson}}(2021)}]{hooshmand}%
  \BibitemOpen
  \bibfield  {author} {\bibinfo {author} {\bibnamefont {{Zahra Hooshmand,
  Jie-Xiang Yu, Hai-Ping Cheng, and Mark R Pederson}}},\ }\bibfield  {title}
  {\bibinfo {title}
  {\href{https://doi.org/10.1103/PhysRevB.104.134411}{Electronic control of
  strong magnetic anisotropy in Co-based single-molecule magnets}},\
  }\href@noop {} {\bibfield  {journal} {\bibinfo  {journal} {Phys. Rev. B}\
  }\textbf {\bibinfo {volume} {104}},\ \bibinfo {pages} {134411} (\bibinfo
  {year} {2021})}\BibitemShut {NoStop}%
\bibitem [{\citenamefont {{Shuanglong Liu, Maher Yazback, James N. Fry,
  Xiao-Guang Zhang, and Hai-Ping Cheng}}(2022)}]{liu}%
  \BibitemOpen
  \bibfield  {author} {\bibinfo {author} {\bibnamefont {{Shuanglong Liu, Maher
  Yazback, James N. Fry, Xiao-Guang Zhang, and Hai-Ping Cheng}}},\ }\bibfield
  {title} {\bibinfo {title}
  {\href{https://doi.org/10.1103/PhysRevB.105.035401}{Single-molecule magnet
  Mn12 on GaAs-supported graphene: Gate field effects from first principles}},\
  }\href@noop {} {\bibfield  {journal} {\bibinfo  {journal} {Phys. Rev. B}\
  }\textbf {\bibinfo {volume} {105}},\ \bibinfo {pages} {035401} (\bibinfo
  {year} {2022})}\BibitemShut {NoStop}%
\bibitem [{\citenamefont {Follmer}\ \emph {et~al.}(2020)\citenamefont
  {Follmer}, \citenamefont {Ribson}, \citenamefont {Oyala}, \citenamefont
  {Chen},\ and\ \citenamefont {Hadt}}]{hadt}%
  \BibitemOpen
  \bibfield  {author} {\bibinfo {author} {\bibfnamefont {A.~H.}\ \bibnamefont
  {Follmer}}, \bibinfo {author} {\bibfnamefont {R.~D.}\ \bibnamefont {Ribson}},
  \bibinfo {author} {\bibfnamefont {P.~H.}\ \bibnamefont {Oyala}}, \bibinfo
  {author} {\bibfnamefont {G.~Y.}\ \bibnamefont {Chen}},\ and\ \bibinfo
  {author} {\bibfnamefont {R.~G.}\ \bibnamefont {Hadt}},\ }\bibfield  {title}
  {\bibinfo {title}
  {\href{https://doi.org/10.1021/acs.jpca.0c07860}{Understanding Covalent
  versus Spin–Orbit Coupling Contributions to Temperature-Dependent Electron
  Spin Relaxation in Cupric and Vanadyl Phthalocyanines}},\ }\href
  {https://doi.org/10.1021/acs.jpca.0c07860} {\bibfield  {journal} {\bibinfo
  {journal} {J. Phys. Chem. A}\ }\textbf {\bibinfo {volume} {124}},\ \bibinfo
  {pages} {9252} (\bibinfo {year} {2020})}\BibitemShut {NoStop}%
\bibitem [{\citenamefont {{M. Zwierzycki, Petr Khomyakov, A.A. Starikov, K.
  Xia, M. Talanana, P.X. Xu, Volodymyr Karpan, I. Marushchenko, I. Turek, E.W.
  Bauer, G. Brocks, Kelly, Paul J.}}(2008)}]{zwierzycki}%
  \BibitemOpen
  \bibfield  {author} {\bibinfo {author} {\bibnamefont {{M. Zwierzycki, Petr
  Khomyakov, A.A. Starikov, K. Xia, M. Talanana, P.X. Xu, Volodymyr Karpan, I.
  Marushchenko, I. Turek, E.W. Bauer, G. Brocks, Kelly, Paul J.}}},\ }\bibfield
   {title} {\bibinfo {title}
  {\href{https://onlinelibrary.wiley.com/doi/10.1002/pssb.200743359}{Calculating
  Scattering Matrices by Wave Function Matching}},\ }\href@noop {} {\bibfield
  {journal} {\bibinfo  {journal} {Phys. Status Solidi B}\ }\textbf {\bibinfo
  {volume} {245}} (\bibinfo {year} {2008})}\BibitemShut {NoStop}%
\bibitem [{\citenamefont {{R. Shankar}}(1994)}]{shankar}%
  \BibitemOpen
  \bibfield  {author} {\bibinfo {author} {\bibnamefont {{R. Shankar}}},\
  }\href@noop {} {\emph {\bibinfo {title} {Principles of Quantum mechanics}}},\
  \bibinfo {edition} {2nd}\ ed.\ (\bibinfo  {publisher} {Plenum Press},\
  \bibinfo {address} {New York},\ \bibinfo {year} {1994})\BibitemShut {NoStop}%
\end{thebibliography}%

\end{document}